\documentclass{pasa}%

\usepackage{rotating}
\usepackage{graphicx}
\usepackage{amssymb} 

\usepackage[final, breaklinks,colorlinks,
   urlcolor=blue,citecolor=blue,linkcolor=blue]{hyperref}

\usepackage{aas_macros}
\def\arcsec   {\hbox{$^{\prime\prime}$}}
\def\degr     {\hbox{$^\circ$}}
\def\arcmin   {\hbox{$^{\prime}$}}
\newcommand{\fig}{Fig.}

\newcommand{\tab}{Table}

\newcommand{\sect}{Section}
\newcommand{\sects}{Sections}

\newcommand{\eqn}{Equation}
\newcommand{\eqns}{Equations}

\newcommand{\Fig}{Fig.}

\newcommand{\Eqn}{Equation}

\title[MWA Phase II IPS DR1]{A census of compact sources at 162MHz: first data release from the MWA Phase II IPS Survey}

\author[J. S. Morgan et al.]{J. S. Morgan,$^{1}$ R. Chhetri,$^{1,2}$ R. Ekers,$^{3,1}$
\affil{$^{1}$International Center for Radio Astronomy Research, Curtin University, GPO Box U1987, Perth, WA 6845, Australia}
\affil{$^{2}$CSIRO Space and Astronomy, P.O. Box 1130, Bentley, WA 6102, Australia}
\affil{$^{3}$CSIRO Space and Astronomy, P.O. Box 76, Epping, NSW 1710, Australia}
}

\jid{PASA}
\doi{10.1017/pas.\the\year.xxx}
\jyear{\the\year}

\hypersetup{draft}

\begin{document}

\begin{frontmatter}
\maketitle

\begin{abstract}
We present a catalogue of over 7000 sources from the GLEAM survey which have significant structure on sub-arcsecond scales at 162\,MHz.
The compact nature of these sources was detected and quantified via their Interplanetary Scintillation (IPS) signature, measured in interferometric images from the Murchison Widefield Array.
The advantage of this approach is that all sufficiently compact sources across the survey area are included down to a well-defined flux density limit.
The survey is based on $\sim$250$\times$ 10-minute observations, and the area covered is somewhat irregular, but the area within 1\,hr$<$RA$<$11\,hr; $-10\degr<$Decl.$<+20\degr$ is covered entirely, and over 85\% of this area has a detection limit for compact structure below 0.2\,Jy.
7\,839 sources clearly showing IPS were detected ($>5\sigma$ confidence), with a further 5\,550 tentative ($>2\sigma$ confidence) detections.
Normalised Scintillation Indices (NSI; a measure of the fraction of flux density coming from a compact component) are reported for these sources.
Robust and informative upper limits on the NSI are reported for a further 31\,081 sources.
This represents the largest survey of compact sources at radio frequencies ever undertaken.
\end{abstract}

\begin{keywords}
Radio continuum: Radio continuum: galaxies -- Scattering -- Sun: heliosphere -- techniques: interferometric -- Techniques: high angular resolution -- Radio continuum: ISM
\end{keywords}
\end{frontmatter}

\section{INTRODUCTION}
\label{sec:intro}
The presence or absence of radio source structure at sub-arcsecond scales allows the separation of more compact structure indicative of present or recent activity (such as cores and hotspots) from the large radio lobes that dominate the source population at low radio frequencies.
Nonetheless, large, unbiased surveys of compact structure have, until recently, been lacking in the literature, due to the technical difficulties of surveying large areas of the sky at very high resolution (i.e. with long baselines), particularly at low frequencies.
As a result, with a handful of exceptions \citep[e.g.][]{2004evn..conf...31P,2014AJ....147...14D} most observing campaigns aimed at identifying compact sources have focused on relatively small fields using the widefield VLBI approach \citep{2001A&A...366L...5G,2011A&A...526A..74M,2013ApJ...768...12M,2018A&A...619A..48R}  and/or have selectively observed sources likely to be compact, based on their flat spectra. \citep[e.g.][]{2002ApJS..141...13B,2015A&A...574A..73M,2016A&A...595A..86J}.

The discovery of Interplanetary Scintillation (IPS) by \citet{Clarke:1964} led \citet{1964Natur.203.1214H} to propose IPS as an alternative method for determining which radio sources have a compact component.
Since IPS arises from interference between radio waves that traverse the solar wind several hundred kilometres apart (for metre wavelengths), coherence is destroyed for sources much larger than 1$\arcsec$, and the scintillation signal will be suppressed.
This led eventually to several catalogues of IPS sources \citep{1987MNRAS.229..589P,1993BASI...21..469B}\footnote{The Ooty catalogue itself was never published, but catalogues from several IPS observatories are available on request. See \citet{2010SoPh..265..309M}.}.

In previous work, IPS was reintroduced as a viable method for identifying compact sources \citep{2018MNRAS.473.2965M} using the most recent generation of low-frequency radio interferometers, in particular the Murchison Widefield Array \citep[MWA;][]{2013PASA...30....7T}.
With a handful of proof-of-concept observations, new compact sources were identified and their properties explored \citep{2018MNRAS.474.4937C,2018MNRAS.479.2318C,2019MNRAS.483.1354S}.
\citet{2022MNRAS.509.2122J} extended this work to show that compactness as measured by MWA IPS observations was correlated with structure measured with GHz VLBI, which probes scales an order of magnitude more compact at much higher frequencies.
A key point is that almost all the sources that we observed were also present in the GLEAM survey \citep{2015PASA...32...25W,2017MNRAS.464.1146H}, which provides detailed flux density measurements in the range 72--231\,MHz.
We also explored the feasibility of an all-sky IPS Survey \citep{2019PASA...36....2M} using the MWA, demonstrating that it would be possible to detect many thousands of sources within 30$\degr$\ of the ecliptic.

Here we present the culmination of several years of effort to generate a catalogue of compact sources over a large area using the IPS technique.
We use data from extended configuration of the Phase-II MWA \citep{2018PASA...35...33W}, which, as discussed by \citet{2019PASA...36...50B}, is superior to the Phase I MWA for IPS observations due to its more uniform ($u, v$) coverage.
Although this IPS survey has already been used in a number of publications \citep{2021PASA...38...49D,2022A&A...658A...2J,2022arXiv220408490B}, using preliminary versions of the catalogue presented here, this is the first release of IPS data to the community from this observing campaign, and the first comprehensive description of the scheduling, observations, data reduction, and synthesis of the catalogue.

An important recent development is the demonstration of the International LOFAR Telescope \citep{2013A&A...556A...2V} to be a viable instrument for conducting wide-field surveys, even while using the international (100\,km--1000\,km) baselines, and thus probing very similar spatial scales to IPS \citep{2022A&A...658A...1M}; indeed \citet{2022NatAs...6..350S} detected a staggering 2483 sources with a compact component in a synthesis image with a resolution of 0.38\arcsec$\times$0.3\arcsec, 6.6 square degrees in extent.
Our survey is highly complementary, since it covers a far wider field of view to a much shallower depth.
In addition, we are observing from the Southern Hemisphere, from the site of the future Low-frequency Aperture Array of the Square Kilometer Array (SKA\_LOW). 
While we are most sensitive when surveying along the ecliptic, much of this area is too far south to be easily observed by LOFAR (though there is, by design, some overlap in this first data release).
Our approach therefore allows us to survey a very large fraction of the sky down to a flux density limit that encompasses a large fraction of GLEAM sources, while avoiding difficulties of low-frequency VLBI imposed by the ionosphere, and the enormous computational cost required to generate very large interferometric images.

This first data release has at least three uses.
Primarily we intend it to be an astrophysical dataset which is useful in its own right, while being highly synergistic with the GLEAM survey which we use as our reference catalogue.
Secondly, the catalogue will be useful in providing compact calibrators for other low-frequency, high-resolution instruments such as LOFAR and the future SKA\_LOW (including the use of the latter as the core of a VLBI array).
Finally, our catalogue will be invaluable for space weather studies using IPS, providing a network of IPS sources (with their baseline scintillation indices) which is unprecedented in its sky density.

The paper is organised as follows.
In \sect~\ref{sec:methods1} we describe comprehensively the process of scheduling observations, selecting the first data release, calibrating and imaging our observations, making image-based measurements of source brightness both in continuum and variability.
In \sect~\ref{sec:methods2} we describe how multiple measurements of each source are synthesised into a single catalogue entry per source.
In \sect~\ref{sec:sensitivity} we discuss the sensitivity of our survey, and discuss issues that have the potential to impact on reliability and completeness.
In \sect~\ref{sec:discussion} we discuss future work.

\section{Methods 1: Scheduling, Calibration, Imaging and Source-finding}
\label{sec:methods1}
The basic approach of using interferometric imaging to make IPS measurements in a single observation is described by \citet{2018MNRAS.473.2965M}.
\citet{2019PASA...36....2M} expand this to performing a survey with multiple observations, and provide some prospective ideas on data reduction and processing, with a focus on MWA IPS data taken in late 2015 through to mid 2016.
 
Since the start of 2019, we have taken a great deal more IPS data, with the MWA in its Phase II extended configuration \citep{2018PASA...35...33W}, which has longer baselines (up to 5.3\,km) and more uniform ($u, v$) coverage than the Phase I MWA.
The potential advantages of the Phase II MWA for IPS studies are set out in detail by \citet{2019PASA...36...50B} \sect~6.2.1. 
In light of these advantages, and the more uniform coverage (mostly due to the automatic scheduling algorithm described in \sect~\ref{sec:scheduling}), we have chosen to use only MWA Phase II data for this data release.

Below we describe our full methodology from scheduling the observations to generating the final catalogue.
Much of this methodology has been previously presented by \citet{2018MNRAS.473.2965M} and \citet{2019PASA...36....2M}.
Where this is the case we have provided a summary and a reference to the relevant section of the other paper.

\subsection{Scheduling}
\label{sec:scheduling}
In order to provide good coverage for both astrophysical and space weather studies, we made daily 10-minute observations of a number of target fields at an elongation from the Sun of approximately 30\degr, at a number of different orientations relative to ecliptic North.
As in previous work, we split the available 30.72\,MHz of instantaneous bandwidth into two equal bands centred on approximately 80\,MHz and 162\,MHz.
Only the upper band has been processed so far.
The principal daily targets were due East and West along the ecliptic (i.e. 90\degr\ and 270\degr\ relative to Ecliptic North) as well as 60\degr, 120\degr, 240\degr, and 300\degr.
At times (depending on the Declination of the Sun, and how much time was available) we added target fields at position angles of 30\degr\ and 330\degr; or at 150\degr, 180\degr\ and 210\degr, all at 30\degr\ solar elongation.
These pointings overlap at the half-power point or closer, therefore providing uniform sensitivity.

As well as maximising the target sensitivity it is also necessary to minimise the response of the instrument at the location of the Sun. 
To facilitate automated, optimal scheduling of these observations we used the model of \citet{2017PASA...34...62S} to pre-calculate beams for all 197 ``sweetspot'' pointings of the MWA at 162\,MHz (the reader is referred to \citet{2019PASA...36....2M} \sect~3.1 for a much more detailed description of MWA beams).

For each day of observations, an optimised pointing and observing time for each target was chosen as follows.
All possible pointings and solar hour angles were exhaustively searched (the latter with a resolution of 1\degr) to find which best match our criteria: the highest target sensitivity with at least 20dB of suppression at the location of the Sun.

Next, a higher-resolution search was carried out to determine the precise 10-minute observing interval over which the Sun is best nulled. Again, this was an exhaustive search, this time with a resolution of 8\,s, since all MWA observations must start and stop at a time when the seconds since midnight is divisible by 8.

To avoid clashes between observations, the targets for the day's observing were scheduled in strict order (those closest to the ecliptic were typically scheduled first).
Code for automated scheduling is available on github\footnote{\url{github.com/johnsmorgan/ips_plan}}. 

Overall, this resulted in 1448 observations scheduled almost every day between 2019-02-04 and 2019-08-18.

\subsection{Array Calibration}
\label{sec:array_cal}
Each observation was downloaded as a measurement set\footnote{\url{asvo.mwatelescope.org}}, and calibrated \citep{2015PASA...32....8O} against a sky model based on the GLEAM survey \citep{2017MNRAS.464.1146H} using publicly available code \citep{2022arXiv220412762H}\footnote{\url{github.com/nhurleywalker/GLEAM-X-pipeline}}, with a few brighter sources from outside the GLEAM survey area characterised on the basis of other radio surveys\footnote{\url{github.com/johnsmorgan/marco}}.
Only baselines between 130\,m and 2600\,m were used for calibration.
The higher cut-off was due to the maximum baseline of (MWA phase I) GLEAM.
The lower cut-off was to avoid contributions from the Sun (which was not included in the sky model due to its variable nature) which is mostly resolved out on baselines $\gtrsim65\lambda$ (see \sect~\ref{sec:imaging} for further information).
These calibration solutions consist of complex gains for each spectral channel for all 4 correlations products (XX, XY, YX, YY), for all 128 tiles.
We then used these calibration solutions to triage our data.
Two metrics were then used to determine the goodness of the calibration. 
The first was the fraction of the calibration solution for which a solution was obtained, the 2nd was the residual of a linear fit of the calibration solution phase as a function of frequency.  
These two metrics were then used to select observations for further analysis.
Broadly speaking, around 88\% of observations met our threshold for further analysis. 
Most observations that did not reach this threshold failed due to known issues with the array at that time. 
Some useful information can likely be gleaned from these observations in the future with more careful calibration and flagging.

\subsection{Selection of Data Release 1}
\begin{figure*}
    \centering
    \includegraphics[width=1.0\textwidth]{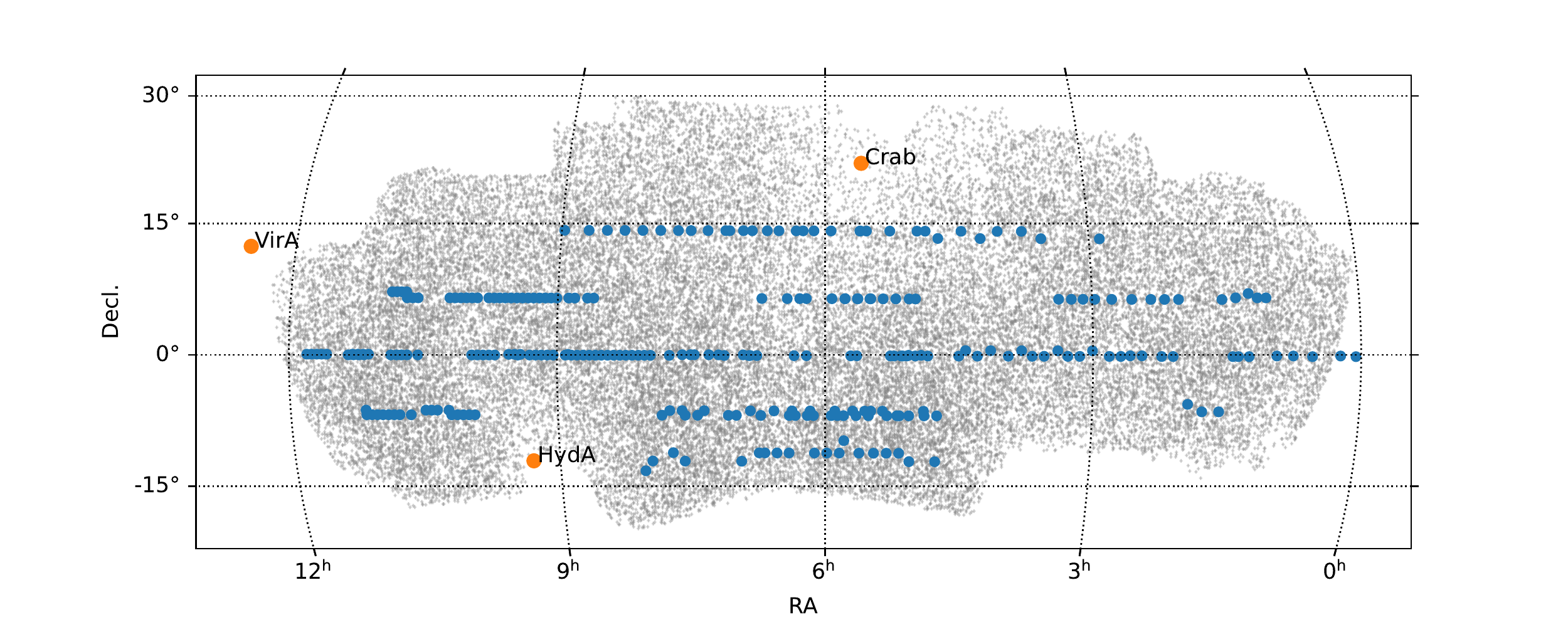}
    \caption{\label{fig:pointings} Pointing centres for 263 observations selected for data release 1 are marked with blue points. The 7839 sources appearing in the final catalogue are plotted in grey to indicate the final coverage (an apparent lack of sources associated with the westernmost pointings is due to the criteria that a source be detected in 5 observations -- see \sect~\ref{sec:gleamxmatch}). Very bright `A-team' sources are also shown.}
\end{figure*}

Observations were chosen to have good overlap with deep infrared surveys and high-resolution radio surveys, to allow a continuation of the work begun by \citet{2019MNRAS.483.1354S}.
We have also focused initially on Northern-hemisphere observations, to facilitate the comparison we have made with International LOFAR \citep{2022A&A...658A...2J}.
These were then supplemented with observations covering the Galactic Plane and Orion region for pulsar searches (Chhetri et al. in prep) and measurements of Galactic scattering (Morgan et al. in prep.). 
For the Easternmost part of the survey area we processed all observations taken over 49 days for two pointings, resulting in a very dense oversampling of the sky.
For the rest of the survey we imaged only a subset of observations, resulting in around half the density (as can be seen in figure~\ref{fig:pointings}).
In all, 263 observations were selected, all observed between 2019-02-21 and 2019-08-18 (close to solar minimum).

\subsection{Imaging}
\label{sec:imaging}
For imaging purposes, measurement sets were generated with the native time  resolution of 0.5\,s, and a spectral resolution of 160\,kHz. 
The outer 160\,kHz channels of each 1.28-MHz coarse channel were flagged.
This level of spectral averaging will cause some degree of bandwidth smearing towards the edge of the beam \citep{2015PASA...32....6O}; but since we are mostly interested in the scintillation index of our sources, anything that effects the numerator (variability) and the denominator (mean brightness) in proportion will cancel out (except for a modest reduction in signal-to-noise due to slight decoherence on the longer baselines).
\textsc{chgcentre}, a companion tool to \textsc{WSClean} \citep{2014MNRAS.444..606O}, was used to adjust the phase centre to the true primary beam maximum, and then to rotate the phases to the minimum-w direction (i.e. close to the zenith).
The latter dramatically reduces the computation required for imaging by reducing the number of w-layers required\footnote{See, e.g., \citealp{1999ASPC..180..383P} for a description of the w-term problem, and \citealp{2014MNRAS.444..606O} for the approach that \textsc{WSClean} uses to solve it.}, as well as ensuring that the PSF is as uniform as possible (in pixel space) across the resulting image.

Each observation was imaged twice: first a continuum image using all 10\,minutes of data with a fairly deep clean was generated (hereafter called the ``standard image'').
This is to provide a means to measure the (mean) flux density of each source of interest (the divisor in the scintillation index).
Secondly, each 0.5\,s timestep of the observation is imaged separately, with only a shallow image-based clean (hereafter called the `snapshot images').
In order to ensure that precisely the same data were used for the dividend and divisor of the scintillation index, we made the decision to use precisely the same imaging image size, pixel size and visibility averaging and weighting for both imaging runs.
This approach is feasible because the extended Phase II MWA has excellent ($u, v$) coverage, and weighting schemes close to natural produce very good images.
An innovation we have introduced since previous work is to subtract the model (that results from the deep cleaning of the standard image) from the visibilities before imaging the snapshots.
This ensures that the cleaning done on the snapshot images is directed towards cleaning (and reducing the sidelobes of) varying sources.
This reduces some artefacts due to very bright continuum sources.

\textsc{WSClean} \citep{2014MNRAS.444..606O,2017MNRAS.471..301O}, and associated tools have been used throughout for calibration, imaging and analysis.
We have found \textsc{WSClean} and associated tools for MWA data reduction \citep{2015PASA...32....8O} to be highly performant, reliable, and flexible.
In particular, the ease of imaging with a model subtracted, and the ability to fine-tune the visibility weighting scheme were extremely valuable for this project.
The latter allowed us to find a weighting scheme which balanced a number of competing factors.
Firstly, our IPS measurements are limited by the system noise (rather than confusion which is often the case \citep[e.g.][]{2015PASA...32...25W}), which favours the use of a weighting scheme that weights all baselines equally (i.e. natural weighting).
Secondly, the Sun remains a contaminant in spite of strong suppression by the beam response of the instrument.
The quiet Sun's power is concentrated in a relatively small number of short baselines, and while this large-scale structure is relatively static and therefore does not change on IPS timescales, it does cause strong ripples in a standard image.

We also wish to minimise the size of our images, both to reduce the computation required for imaging, and the storage required for the snapshot images.
Inclusion of the longest MWA baselines would necessitate a small pixel size and therefore much larger images than used in our MWA phase-I pilot studies.

\begin{figure*}
    \centering
    \includegraphics[width=0.75\textwidth]{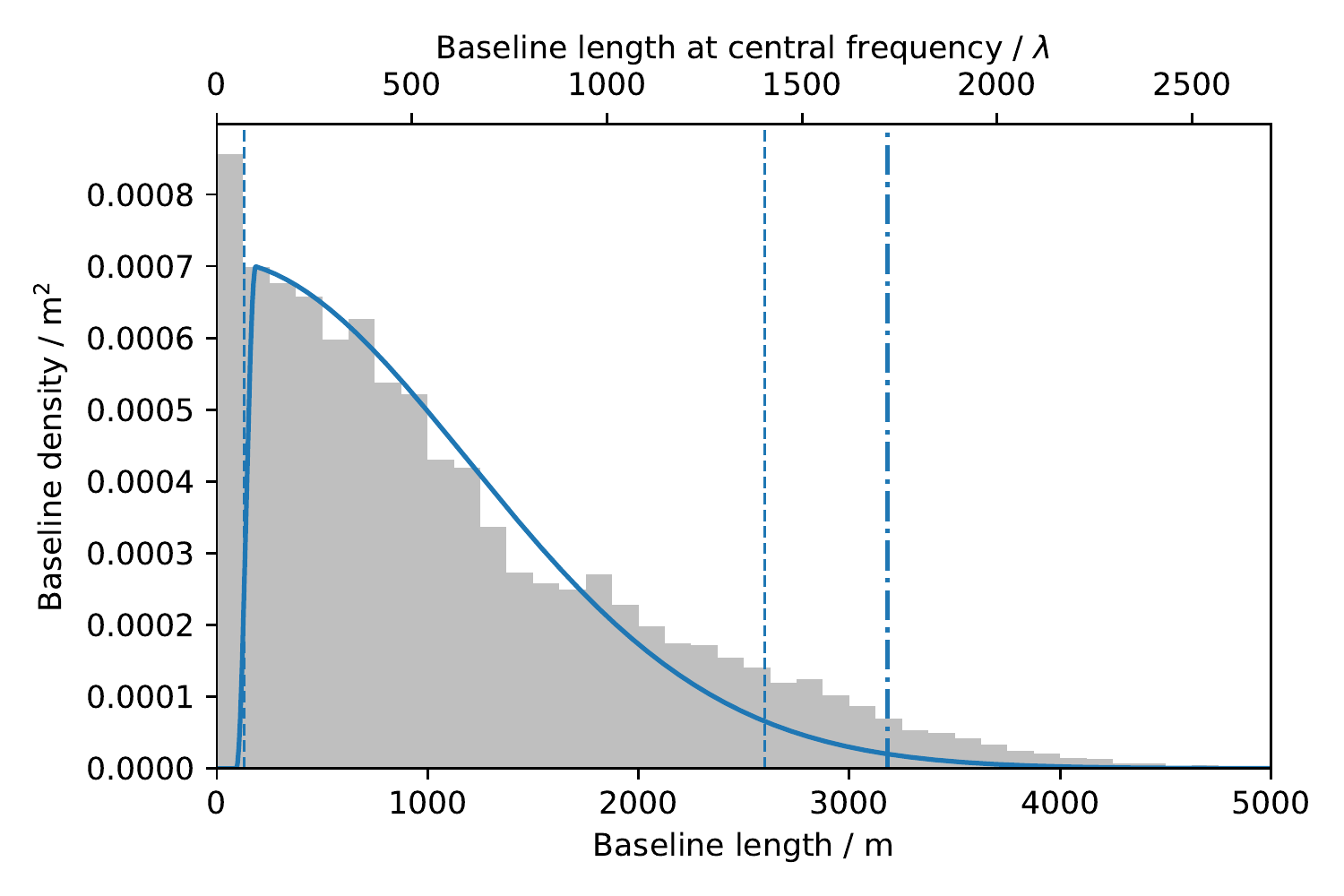}
    \caption{\label{fig:weighting}The solid blue line shows the weighting scheme used, in arbitrary units. This scheme consists of zero weight for all baselines$<50\lambda$ (where $\lambda$ is taken to be 1.85\,m); a Tukey taper from $50\lambda$--$100\lambda$, and a Gaussian taper equivalent to 2\arcmin\ FWHM in the image plane. The vertical dashed lines delimit the range of baseline lengths used for calibration (see \sect~\ref{sec:array_cal}). The dash-dotted line is the Nyquist limit imposed by the 1\arcmin\ pixel size in the image plane. The grey bars indicate the \emph{density} of baselines in each annulus of the ($u, v$) plane.}
\end{figure*}
Our compromise weighting scheme is summarised in \fig~\ref{fig:weighting}.
Uniform weighting is applied to the data by \textsc{WSClean} by default.
Baselines$<50\lambda$ were then discarded completely.
A Tukey taper \citep{Harris:1978} was applied to baselines of length $50\lambda<B\leq100\lambda$.
Finally, a Gaussian taper was applied to the data, equivalent to an image-plane FWHM of 2\arcmin.

As shown in \fig~\ref{fig:weighting}, the Gaussian taper follows the baseline density as a function of baseline length, so its main effect is to reverse the uniform weighting and restore a more natural weighting scheme, albeit with strongly reduced weighting on the longest baselines to bring the resolution in line with the large pixel size.
The large pixel size also excludes approximately 7\% of baselines from being gridded.
Nonetheless, tests showed that with these parameters the image noise (as measured by subtracting two consecutive snapshot images and measuring the standard deviation) was only approximately 10\% higher than natural weighting.

The 2\arcmin\ Gaussian taper also has the advantage that our resolution is well-matched to GLEAM \citep{2017MNRAS.464.1146H} at the appropriate frequency, making it easy to perform absolute flux density corrections.

For the snapshot images only, we also limited the number of w-layers to 24 (the number of CPU cores on the machine we were using).
This reduced the computation time further, and in spite of the \textsc{WSClean} ``suggested'' number of w-layers being up to an order of magnitude higher, this did not appear to cause any problematic effects.
All snapshots excluding the first 4\,s and the last 16\,s were imaged \citep[c.f.][]{2022PASA...39....3T}, leaving 1152 timesteps per 10-minute observation.

Both XX and YY correlation products were imaged separately, and the resulting snapshot images (along with XX and YY standard images and beams) were stored in a single HDF5 precisely as described by \citet{2018MNRAS.473.2965M}, Appendix I.
Calibration scaling factors as described below were subsequently stored in the same file, and these are applied on-the-fly when any relevant data are extracted for subsequent analysis.

\subsection{Post-imaging calibration}
\label{sec:post_cal}
In order to both combine both polarisations into ``pseudo Stokes I'' ($I_I$) with optimal weightings, and apply an absolute calibration factor, we use \citet{1996A&AS..120..375S}, \eqn~1, in a very slightly modified form:
\begin{equation}
	\label{eqn:sault}
    I_I(l) = \frac{\Sigma_p A(p) B(l, p) I(l, p) / \sigma^2(p)}{\Sigma_p A^2(p) B^2(l, p) / \sigma^2(p)}
\end{equation}
Here the sum is over the two polarisations $p$ (XX and YY)\footnote{NB: X and Y here are the instrumental polarisations corresponding to ground-based dipoles oriented East-West and North-South respectively. This does not accord with the IAU/IEEE definition of X and Y \citep[][and references therein]{1996A&AS..117..161H} and the two are not necessarily orthogonal. However, this is unimportant for our purpose of combining these measurements while minimising noise.}.
We separate the beam into an absolute correction factor $A$ per polarisation (which, like the variance $\sigma^2$ is assumed to be direction-independent), and a beam, $B(l)$ which is direction-dependent (denoted by it being a function of $l$), and generated for each pixel using the model of \citet{2017PASA...34...62S}.
For these, for efficiency, we pre-calculated all-sky beams and interpolated these as required using software available on github\footnote{\url{github.com/johnsmorgan/mwa_pb_lookup}} and archived on Zenodo \citep{2021zndo...5083990M}.

After imaging, \textsc{Aegean 2.2.0} \citep{2012MNRAS.422.1812H,2018PASA...35...11H} was used to generate a source catalogue for the standard image for XX and YY separately.
After applying primary beam corrections, selecting a subset of bright, unresolved sources, and comparing with the relevant measurements from GLEAM \citep{2017MNRAS.464.1146H} these could be used to determine the absolute calibration factor $A$ for each polarisation.
A small variability image \citep{2018MNRAS.473.2965M} for the central quarter of the full image was then generated from the snapshots for each polarisation for the purpose of measuring $\sigma$, which was taken to be the median of this image.
Note that the direction-independent $\sigma\left(p\right)$ values are only used for combining the two polarisations.
When the noise is measured for a particular source, either the standard or variability images, as described below, a local measurement of Root-Mean-Square (RMS) is used.

Once the factors $A(p)$ and $\sigma(p)$ were calculated, it was possible to combine both polarisations of the standard image together using \eqn~\ref{eqn:sault}.
A variability image was then constructed using the timeseries data, with polarisations combined in the same manner, exactly as described by \citet{2018MNRAS.473.2965M}, \sect~2.3.
Essentially, the variability image is produced by taking the timeseries corresponding to each pixel of the image, applying a filter with a bandpass of 0.1\,Hz--0.4\,Hz, and then computing the RMS.
The filter emphasises the frequency range where the IPS signal is strongest, while filtering out almost all ionospheric scintillation \citep{2022arXiv220713252W}.
Additionally, for this survey, we found that by applying a Tukey window \citep{Harris:1978} to the timeseries before filtering we were able to vastly reduce the number of spurious detections, probably due to instrumental effects near the start and end of observations.

\subsection{Standard image and variability image characteristics}
\label{sec:images}
The result of the process at this point was a single standard image and a single variability image for each observation; these were the only data products that were carried forward for further analysis. 
\textsc{BANE} \citep[part of the \textsc{AegeanTools} suite;][]{2018PASA...35...11H} was used to produce `background' and `rms' images.
MWA continuum images are confusion-limited at a level well above the thermal noise, and the noise tends to be higher around bright sources.
The noise in variability images are generally thermal noise dominated, since the number of variable sources is lower than the total number of sources (and the scintillating flux density is less than the mean flux density).
Sidelobes are only seen around the very brightest sources, and then only occasionally.

Furthermore, since the variability image measures standard deviation, all pixels have a positive value.
The `background' value, is the thermal noise level which is present whether a pixel corresponds to variable source or not.

The `rms' image measures the spatial deviation of the background about its mean.
If the lightcurves consisted of Gaussian random noise, the RMS would be $\chi$ distributed, with 564 degrees of freedom (d.o.f.; after our application of a low-pass filter with a timescale of 1\,s and Tukey window with the taper covering 4\% of the timeseries).
In practice, we observe a ratio of 32.25 between the background and RMS of the variability images, and this is remarkably consistent across all observations for all points in the image.
This ratio implies a $\chi$ distribution with $\sim$520 d.o.f.
The $\chi$ distribution converges towards a normal distribution with increasing degrees of freedom (more rapidly than the better known $\chi^2$ distribution), justifying our assumption of Gaussian noise in the variability image (though we assume a $\chi$ distribution with 520 d.o.f. when analysing our false detection rate; see \sect~\ref{sec:false}).

\subsection{Source finding and characterisation}
Most MWA continuum surveys \citep[e.g. MWACS, GLEAM;][]{2014PASA...31...45H, 2017MNRAS.464.1146H} mosaic together multiple observations in order to increase ($u, v$) coverage and sensitivity before source-finding \citep[see][for a counterexample]{2016MNRAS.461.4151C}.
This allows the detection of sources that would fall below the level of significance in a single observation.
We take a different approach here, and source-find separately for each observation.
This is primarily due to the detection limit in a variability image only falling with the fourth root of time, meaning that the sensitivity increase from combining observations is fairly modest (40 2-$\sigma$ observations would be required to match the significance of a single 5-$\sigma$ detection).

For us, measuring at least a subset of sources in each observation independently is vital in order to determine how the scintillation index depends on the solar latitude of the piercepoint \citep[i.e. the latitude of the point on the Sun beneath the point of closest approach of the line of sight to the Sun; see \sect~\ref{sec:mano} below and also][\sect~2.4]{2018MNRAS.473.2965M}, as well as checking for drastic space weather variations in a particular part of the sky in a particular observation.
We know exactly to what extent (if at all) each individual observation contributes to a source's catalogued properties, making it easier to exclude particular problematic measurements on a case-by-case basis.

We therefore, used \textsc{Aegean} to catalogue all sources detected at 5-$\sigma$ in all variability and standard images.
To capture information on sources scintillating below 5-$\sigma$, we also made a measurement in the variability image at the location of each detection in the standard image (there were typically an order of magnitude more 5-$\sigma$ detections in the standard image than the variability image).
As will be explained in \sect~\ref{sec:nsi}, this is vital in order to make an unbiased estimate of the compactness of our sources.
These measurements were derived from a simple 2D cubic interpolation of the 3$\times$3 nearest pixels to the continuum position in the variability image, and its corresponding `background' and `rms' images.

Correction of source positions for ionospheric refractive shifts was carried out following \citet{2018MNRAS.473.2965M}, \sect~3.2.1: i.e. the offsets of a subset of bright, isolated sources was used to construct a vector field using a radial basis function technique, and this was used to calculate corrected source coordinates for each detection.
As before, we estimate that the typical error in position is well below 1\arcmin\ after this correction.


\subsection{Consolidation of measurements}
\label{sec:gleamxmatch}
The analysis so far has treated each observation entirely independently.
We now organised our data so that measurements are grouped by source.
At this stage we also rejected any measurements that do not meet the following  criteria for inclusion.
We used GLEAM as our reference catalogue, and only sources in GLEAM are in our catalogue (this excludes very few IPS sources, see \sect~\ref{sec:false} for further details).
For each observation we first identified all GLEAM sources which lay within the quarter-power beamwidth of the primary beam.
For each of these sources we then determined whether a continuum or variability detection had indeed been made for that source (within a match radius of 1\arcmin).
This resulted in approximately 2.5 million detections.
Any variability detections which did not match with a GLEAM source (regardless of whether they had a counterpart in the continuum image) were stored for later analysis of false detections (see \sect~\ref{sec:false}).

Measurements outside the elongation range 20\degr\ to 40\degr\ were discarded
and we further specified that there should be 5 continuum detections of our sources, since we found that with fewer measurements, a discrepant measurement (due to, e.g., a space weather fluctuation) could dominate and cause biased results.
We also required at least one measurement within the half power point of the primary beam.
After these cuts there remained 748\,321 detections of 42\,838 individual GLEAM sources, each with between 5 and 57 measurements.

Note the two possible ways that variability image measurement may be made: either by a direct detection by source-finding in the variability image, or indirectly by measurement at the location of the continuum detection. 
Strong variability detections will have both, the remainder will only have the latter.
For those with both, the direct detection value was used; otherwise the indirect detection was used.

\section{Methods 2: Synthesis of the Catalogue}
\label{sec:methods2}
Following \citet{2018MNRAS.474.4937C}, we quantify the IPS of each of our sources with the Normalised Scintillation Index (NSI), 
The normalisation is with respect to the Sun--source geometry: the scintillation index decreases with increasing distance from the Sun as the solar wind expands into 3D space.
Additionally, the polar solar wind tends to be more diffuse than that emanating from the Sun's equatorial regions.
We follow \citet{1993SoPh..148..153M} in assuming an elliptical form for the contour of equal scintillation index around the Sun, with the minor axis at the poles of the Sun and the major axis at the equator of the Sun (see \sect~\ref{sec:mano}).
The result of this normalisation is that the NSI is zero for a source that shows no IPS, and unity for a source that scintillates like a point source.

In this section, we describe how we determine the NSI of each sources given our data, how we classify our sources as detections or upper limits, how we determine the error on the NSI, and how we determine the sensitivity at the location of each of our sources.

While the mapping of an observed scintillation index to the NSI is straightforward, the process of deriving the scintillation index from the observed fractional variance while avoiding bias and quantifying the error is necessarily complex.
The complication arises from the fact that we are measuring variance due to scintillation in the presence of variance due to thermal noise.
Both types of variance are stochastic processes (albeit with different timescales).

The concept of debiasing the variability index to remove the effects of random noise is not new \citep{2005ApJ...618..108B}; and in previous work we noted that this leads to asymmetric errors \citep{2018MNRAS.473.2965M}.  
However, in the present work, the use of multiple observations of each source, each with different sensitivities, moves us into an unusual domain where the variance of the signal we are looking for is often dwarfed by the variance due to thermal noise.

\subsection{Normalisation of the Scintillation Index}
\label{sec:mano}
The scintillation index of a point source is given by an empirical function of the form
\begin{equation}
    \label{eqn:mano}
    m \propto \lambda \left(e \sin{\epsilon}\right)^{-b} ,
\end{equation}
(\citealp{1993SoPh..148..153M}; see also \citealp{2019PASA...36....2M} \eqns~6\&7)
where $\lambda$ is the observing wavelength, $e$ is the elliptical term, which depends on the solar latitude below the piercepoint and the ellipticity, and $\epsilon$ is the solar elongation.
While the constant of proportionality, the ellipticity and $b$ may be expected to vary during the solar cycle, and from one solar cycle to the next, we find no strong evidence to vary these parameters from their respective standard values of 0.06, 1.5, and 1.6.
\citet{2022A&A...658A...2J} found that IPS estimates of compact flux density were around 20\% lower than LOFAR estimates, which might justify applying a uniform amplitude scaling.
However a new comparison with the most recent version of the catalogue, suggests that the discrepancy may be closer to 10\%.
The sources in common with LOFAR are the Northernmost of our range and very far South for LOFAR.
For now, we do not apply any overall correction, though we may revisit this in a future data release.

We found a major:minor axis ratio of the contour of constant scintillation index of 1.5 to be a good fit to our data,
which is a typical value for solar minimum as found by \citet{1993SoPh..148..153M}.

\subsection{Determination of NSI}
\label{sec:nsi}
For the case of a single observation, the process required to infer the Normalised Scintillation Index for each source above a certain detection threshold is described fully by \citet{2018MNRAS.473.2965M}.
Briefly we first determine $\Delta S$, the scintillating flux density, from the variability image as follows 
\begin{equation}
    \label{eqn:p_to_ds}
    \Delta S = \sqrt{P^2 - \mu^2}
\end{equation}
\citep[][\eqn~4]{2018MNRAS.473.2965M} where $P$ is the pixel value at the location of the source (as determined by fitting in the image plane), and $\mu$ is the background level of the variability image (i.e. the variability due to thermal noise).
Note that neither $\Delta S$ nor $S$ (the flux density measurement in the continuum image) are absolutely flux density calibrated. 
Any errors (due to e.g. a primary beam model error) will cancel in the scintillation index $\Delta S/S$.
Consequently, the current analysis of our data is insensitive to, and largely unaffected by longer-term variability such as intrinsic variability or refractive interstellar scintillation, which in any case is relatively rare at low frequencies \citep{2019MNRAS.482.2484B}.
Where an absolute flux density measurement is required, such for determining sensitivity (\sect~\ref{sec:sensitivity}), we assume that $S$ is equal to the GLEAM flux density at the appropriate frequency (hereafter $S_\textrm{162}$).

We reiterate that in the absence of scintillation, $P-\mu$ has a Gaussian distribution with an expected value of zero, and a variance of $\sigma^2$ (where $\sigma$ is the spatial RMS in the variability image).
In contrast, while there is a monotonic relationship between $\Delta S$ and $P$, it is non-linear.
The errors on $\Delta S$ are non-Gaussian, and $\Delta S$ is not real unless $P>\mu$.

We now wish to determine the NSI on the basis of multiple measurements.
If we were to restrict ourselves to measurements where $P -\mu > 5\sigma$ as in previous work, this would introduce a positive bias.
Even with the $5\sigma$ threshold relaxed, $P-\mu$ can be negative in the presence of noise, and these values do not map to a real-valued $\Delta S$ or NSI.

We have devised a scheme to determine the NSI that best fits \emph{all} our measurements of variability without any need to remove negative or low-S/N measurements.
First, we define the variability S/N $\rho_\textrm{var}$:
\begin{equation}
    \label{eqn:rho_var}
    \rho_\textrm{var} = \frac{P-\mu}{\sigma} .
\end{equation}
This is a `standardised' statistic: in the absence of a scintillation signal it has zero mean and unit variance.
For a particular $\textrm{NSI}$, the scintillating flux density that a source \emph{would} have is simply
\begin{equation}
    \Delta S\left(\textrm{NSI}\right) = S \cdot m_\textrm{pt} \cdot \textit{NSI} ,
\end{equation}
where $m_\textrm{pt}$ is the scintillation index a point source would have (a function of the Sun/source geometry; see \sect~\ref{sec:mano}).
We can then determine $\rho\left(\textrm{NSI}\right)$, the S/N the variability detection \emph{would} have for a given NSI:
\begin{equation}
    \label{eqn:rho_to_ds}
    \rho\left(\textrm{NSI}\right) = \frac{\sqrt{\Delta S\left(\textrm{NSI}\right)^2+\left(32.25\sigma\right)^2}-\left(32.25\sigma\right)}{\sigma} .
\end{equation}
32.25 is the ratio between $\mu$ and $\sigma$ in the variability image (see \sect~\ref{sec:images}), 

We then determine (by iterative fitting) $\textrm{NSI}_\textrm{fit}$, the NSI that minimises the residual sum of squares:
\begin{equation}
    \label{eqn:s}
    \textrm{RSS}\left(\textrm{NSI}\right) = \frac{\sum_i w_i\left(\rho\left(\textrm{NSI}\right)_i - \rho_{\textrm{var}, i}\right)^2}{\sum w_i} ,
\end{equation}
where the sum is taken over all observations and $w_i$ is the weight of the $i$th observation.

For the weights $w_i$ we use the S/N using the continuum flux density $S$ as the numerator and the noise in the \emph{variability} image as the denominator.
This weights our measurements by inverse thermal noise\footnote{This is a compromise between uniform weighting and inverse variance weighting. In addition, an arguable choice would be to further scale the weights by $m_\textrm{pt}$.
This might marginally increase sensitivity by up-weighting measurements where the scintillating flux density is expected to be higher.
However, observations close to the Sun also move closer to the strong scintillation regime where the expected relationship between solar elongation and $m_\textrm{pt}$ may not hold. On the other hand, our chosen scheme weights observations close to the centre of the primary beam highest, moving the effective elongation of our NSIs towards 30\degr, making for a more uniform dataset.}.
\begin{figure*}
    \centering
    \includegraphics[width=0.45\textwidth]{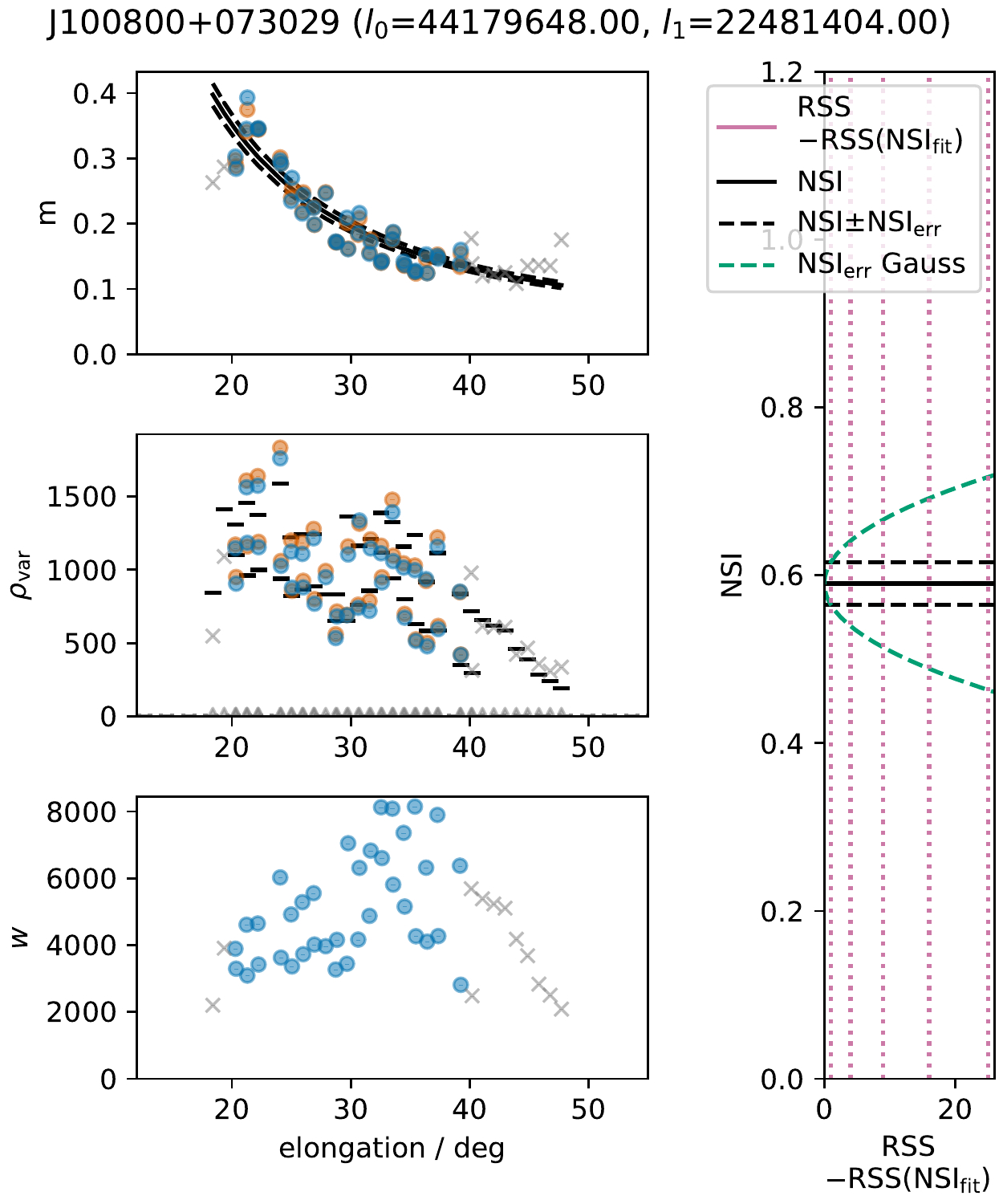} 
    \includegraphics[width=0.45\textwidth]{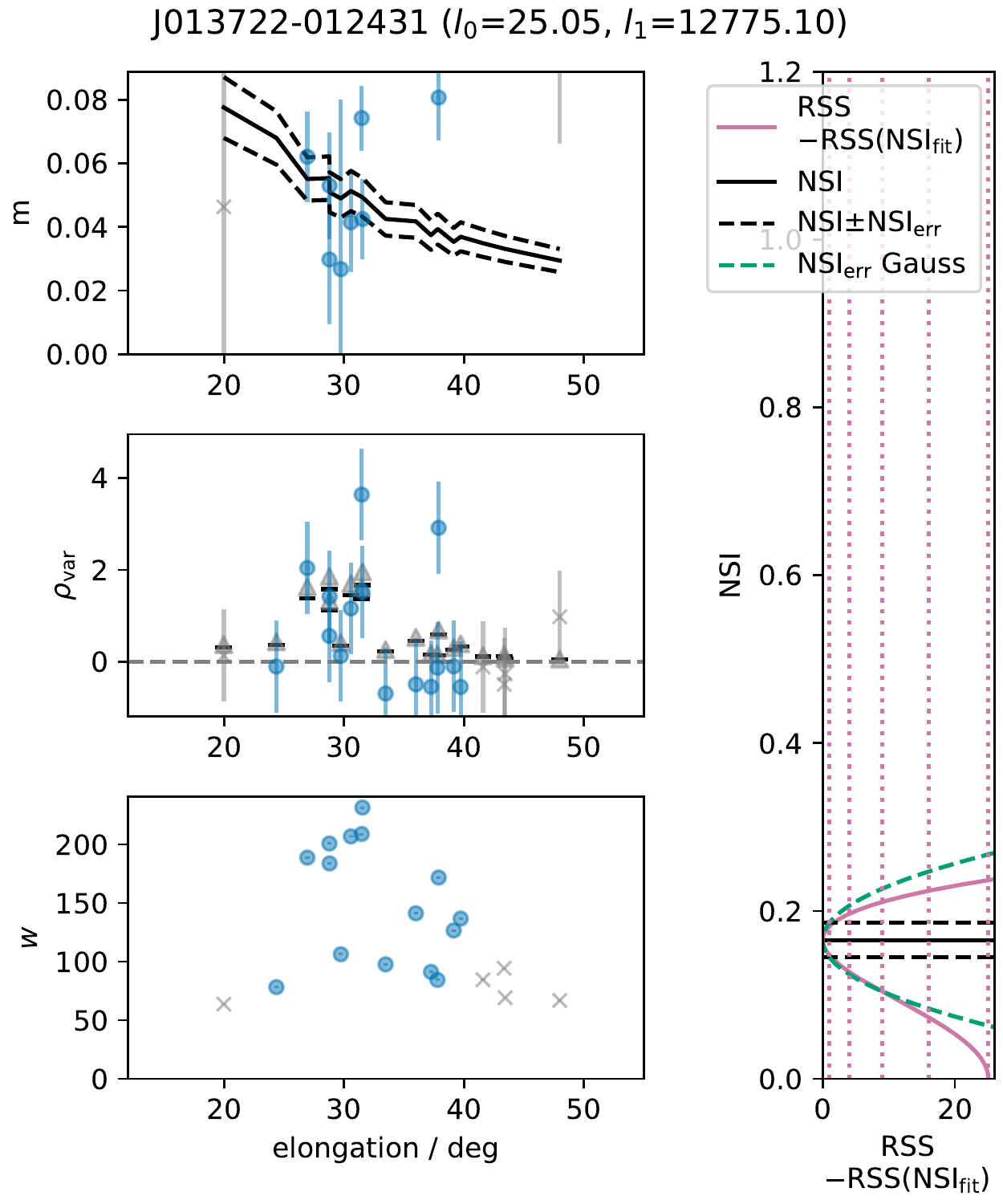} \\ 
    \includegraphics[width=0.45\textwidth]{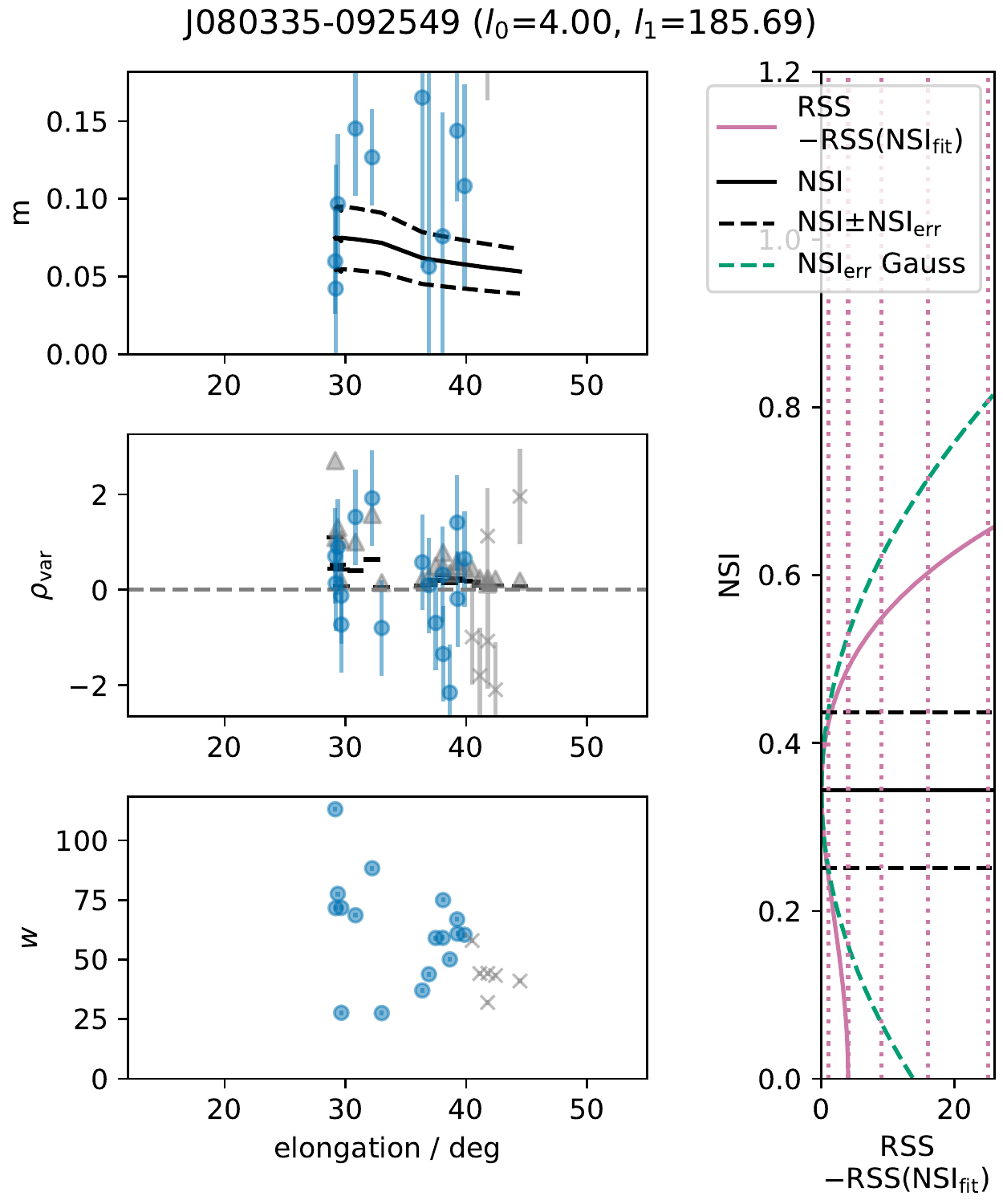} 
    \includegraphics[width=0.45\textwidth]{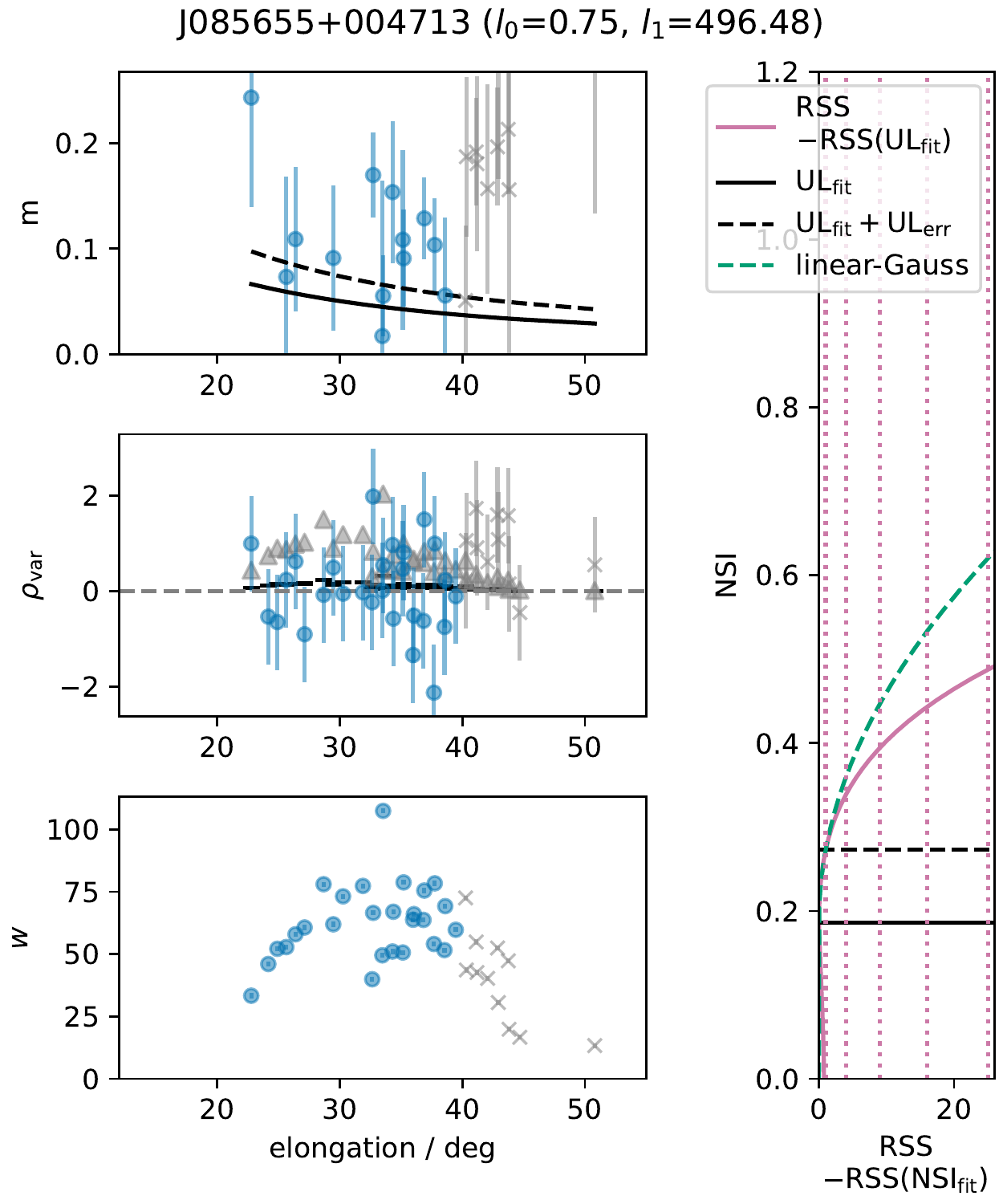} 
    \caption{\label{fig:basic_el}Figures summarising detections and NSI for a high-S/N source (upper left); a 5-sigma source (upper right); a 2-sigma marginal detection (bottom left) and a source for which there is a robust upper limit (bottom right). In each case, top panel shows scintillation index $m$ (\eqn~\ref{eqn:mano}) as a function of elongation, 2nd panels from the top show $\rho_\textrm{var}$ (\eqn~\ref{eqn:rho_var}) and bottom panel shows the weights (\eqn~\ref{eqn:rho_var}). In the top two panels, orange points indicate direct detections, blue indicates indirect detections as defined in \sect~\ref{sec:nsi}. In the top 2 panels, black lines indicate the expected value for the NSI determined by the fit. In the middle panel the grey triangles show the expected value for$\textrm{NSI}_{\textrm{lim}}$ (\eqn~\ref{eqn:ss}). The vertical panels on the right of each figure show the RSS statistic as a function of NSI (\eqn~\ref{eqn:s}). Also plotted is the RSS that would be expected for Gaussian errors given by $\textrm{NSI}_\textrm{err}$ (\sect~\ref{sec:nsi_err}). For the upper-limit source, \eqn~\ref{eqn:ul_logpdf} is used (\sect~\ref{sec:upper_limits}).}
\end{figure*}
This fitting process is summarised in \fig~\ref{fig:basic_el}, which illustrates the relationship between scintillation index, NSI, $\rho_\textrm{var}$, and $w$, as well as showing \textrm{RSS} as a function of NSI.

For purely practical purposes we allow negative NSIs (by giving $\rho\left(\textrm{NSI}\right)$ the sign of the NSI), which are non-physical, but provide a best fit for sources where $\sum\rho_\textrm{var}$ is negative.
The existence of sources with negative NSIs may be indicative of the reliability of the catalogue, so we report their number in \sect~\ref{sec:reliability}.
However, for the final catalogue (and in particular for the detection statistics presented in the next section), $\textrm{NSI}_\textrm{fit}$ is set to zero for all sources with a best-fit NSI$<0$.

\subsection{Classification of detection/non-detection}
\label{sec:class}
Next we use $\textrm{RSS}\left(\textrm{NSI}\right)$ to generate simple statistics to determine the significance of the detection of IPS for each source.
IPS is deemed to have been detected if 
\begin{equation}
    \label{eqn:l_0}
    l_0 = \textrm{RSS}\left(0\right) - \textrm{RSS}\left(\textrm{NSI}_\textrm{fit}\right)>25.
\end{equation}
$l_0$ is twice the log-likelihood ratio between an NSI of $\textrm{NSI}_\textrm{fit}$ and an NSI of zero.
In the signal-absent case (and in the asymptotic limit of a large number of datapoints), $l_0$ is $\chi^2$ distributed with a single degree of freedom (since we have only one free parameter in our fit).
As is well known, the range over which $\textrm{RSS}\left(\textrm{NSI}\right)$ is within 1 of its minimum value is the 68\% (i.e. $1\sigma$) confidence interval.
Due the quadratic form of a Gaussian log likelihood, $l_0=25$ is roughly equivalent to a $5\sigma$ detection \citep{2006smep.book.....J}, 25 being the $l_0$ of a single 5$\sigma$ measurement.

We use a similar statistic, $l_1$, with a slightly less stringent threshold, to determine if we can place an informative upper limit on a source's NSI (for those sources with $\textrm{NSI}_\textrm{fit}<1$):
\begin{equation}
    \label{eqn:l_1}
    l_1 = \textrm{RSS}\left(1\right) - \textrm{RSS}\left(\textrm{NSI}_\textrm{fit}\right)>9.
\end{equation}

Finally, for sensitivity calculations, it is useful to estimate, for each source, the NSI at which the source would only just be detected.
We estimate this by finding the $\textrm{NSI}_\textrm{lim}$ which satisfies the following equation
\begin{equation}
    \label{eqn:ss}
    \frac{\sum w_i \rho\left(\textrm{NSI}_\textrm{lim}\right)_i^2}{\sum w_i} = 25,
\end{equation}
where $\rho\left(\textrm{NSI}\right)$ is as given in \eqn~\ref{eqn:rho_to_ds}.
This is the NSI at which the source would reach our detection threshold in the absence of scatter.

\begin{figure}
    \centering
    \includegraphics[width=1.0\columnwidth]{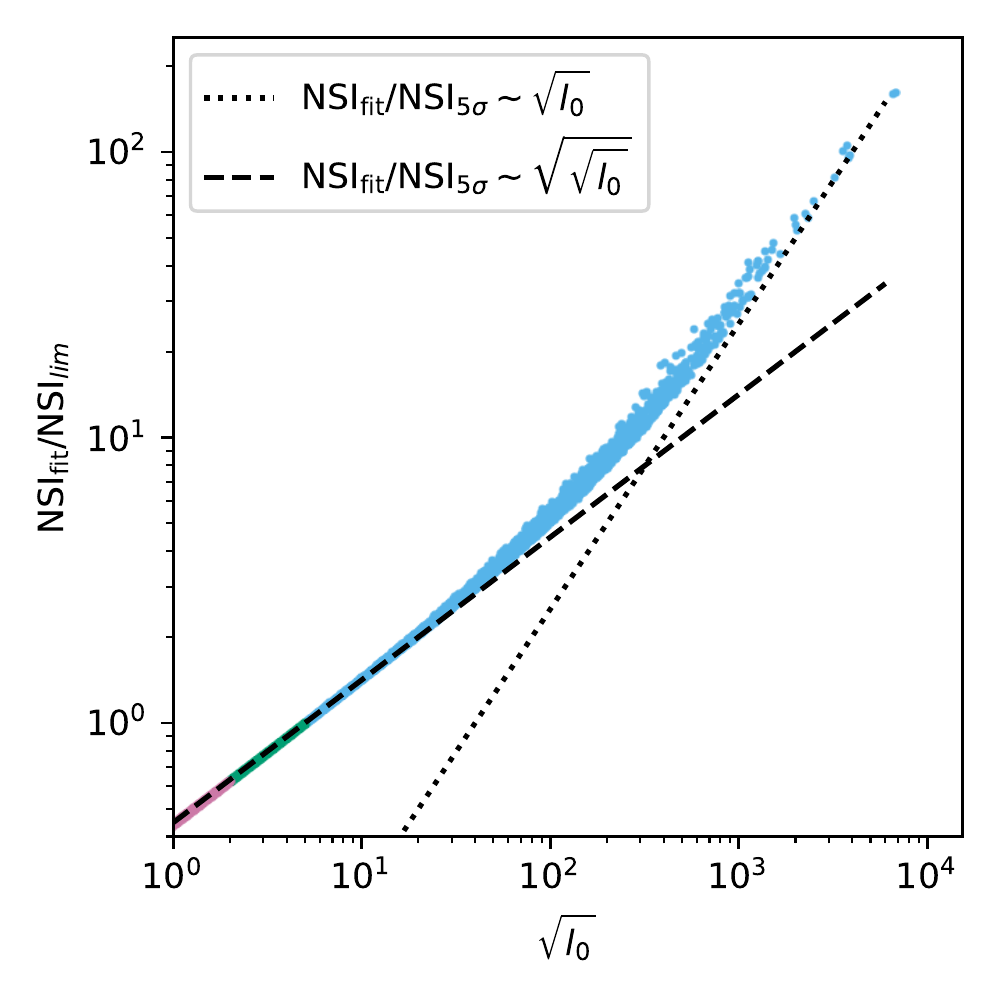}
    \caption{\label{fig:nsi_snr}$\sqrt{l_0}$ (a S/N-like quantity) against the ratio $\textrm{NSI}_\textrm{fit}/\textrm{NSI}_{5\sigma}$. Blue points are detections, green points are marginal detections, pink points are sources with upper limits only. Two trend lines are shown to illustrate that in the weak signal limit, the S/N increases only with the square root of the NSI, and only for the strong detections is there a linear relationship}
\end{figure}

\Fig~\ref{fig:nsi_snr} plots $\sqrt{l_0}$ (essentially the S/N of $\textrm{NSI}_\textrm{fit}$) against the ratio $\textrm{NSI}_\textrm{fit}/\textrm{NSI}_\textrm{lim}$.
The scatter at higher S/N is the effect of space weather (see \sect~\ref{sec:nsi_err}).

Note that in addition to $\textrm{NSI}_\textrm{lim}$ we also calculate $\textrm{NSI}_\textrm{5lim}$.
This is identical to $\textrm{NSI}_\textrm{lim}$ except that only the 5 highest weighted measurements are used.
The significance of 5 is that 5$\times5\sigma$ continuum measurements are required in order for a source to be included in a catalogue, so when multiplied by the flux density of the source to give $S_\textrm{5lim}$, this statistic measures the flux density of the weakest detectable NSI=1 source.
See \sect~\ref{sec:sensitivity_cont_var} for further details.

\subsection{Error on the NSI for detected sources}
\label{sec:nsi_err}
Recall that $\textrm{NSI}_\textrm{fit}$ is our estimate of the NSI of each source.
We take the (1$\sigma$) error due to thermal noise to be $\left(\textrm{NSI}_{\textrm{upper}}-\textrm{NSI}_\textrm{lower}\right)/2$, where
\begin{flalign}
    \label{eqn:nsi_err}
    \textrm{RSS}\left(\textrm{NSI}_\textrm{lower}\right) - \textrm{RSS}\left(\textrm{NSI}_\textrm{fit}\right)= 1; \textrm{NSI}_{\textrm{lower}}<\textrm{NSI}_\textrm{fit}\nonumber &\\
    \textrm{RSS}\left(\textrm{NSI}_\textrm{upper}\right) - \textrm{RSS}\left(\textrm{NSI}_\textrm{fit}\right)= 1; \textrm{NSI}_{\textrm{upper}}>\textrm{NSI}_\textrm{fit}
\end{flalign}
(i.e. half the range in NSI over which RSS increases by $<1$).
$\textrm{NSI}_\textrm{lower}$ and $\textrm{NSI}_\textrm{upper}$ are calculated automatically using the \textsc{minos} routine of \textsc{iminuit} \citep{1975CoPhC..10..343J,Venson:1988,2022zndo...3949207D}, which is used throughout for iterative fitting.

There is an additional error due to stochastic changes in the solar wind, as well as a much smaller uncertainty arising from the stochastic nature of scintillation itself \citep[see][\eqn~7]{2018MNRAS.473.2965M}.
To quantify the combined effect of these, we chose a sample of 120\,000 IPS measurements with sufficient S/N that thermal noise is negligible.
The ``$g$''-factor -- the observed scintillation index compared to expected (from $\textrm{NSI}_\textrm{fit}$) -- was then calculated for each.
The distribution of $g$ values had a standard deviation very close to 25\%.

This variance will be amplified by the non-uniform weighting of measurements used when determining $\textrm{NSI}_\textrm{fit}$. 
To mitigate this, we determine the error due to space weather by dividing 25\% by the square root of the ``effective sample size'' \citep{Kish:1965} given by
\begin{equation}
    \label{eqn:n_eff}
    N_{\textrm{eff}} = \frac{\left(\sum w_i\right)^2}{\sum\left(w_i\right)^2} .
\end{equation}

\begin{figure}
    \centering
    \includegraphics[width=1.0\columnwidth]{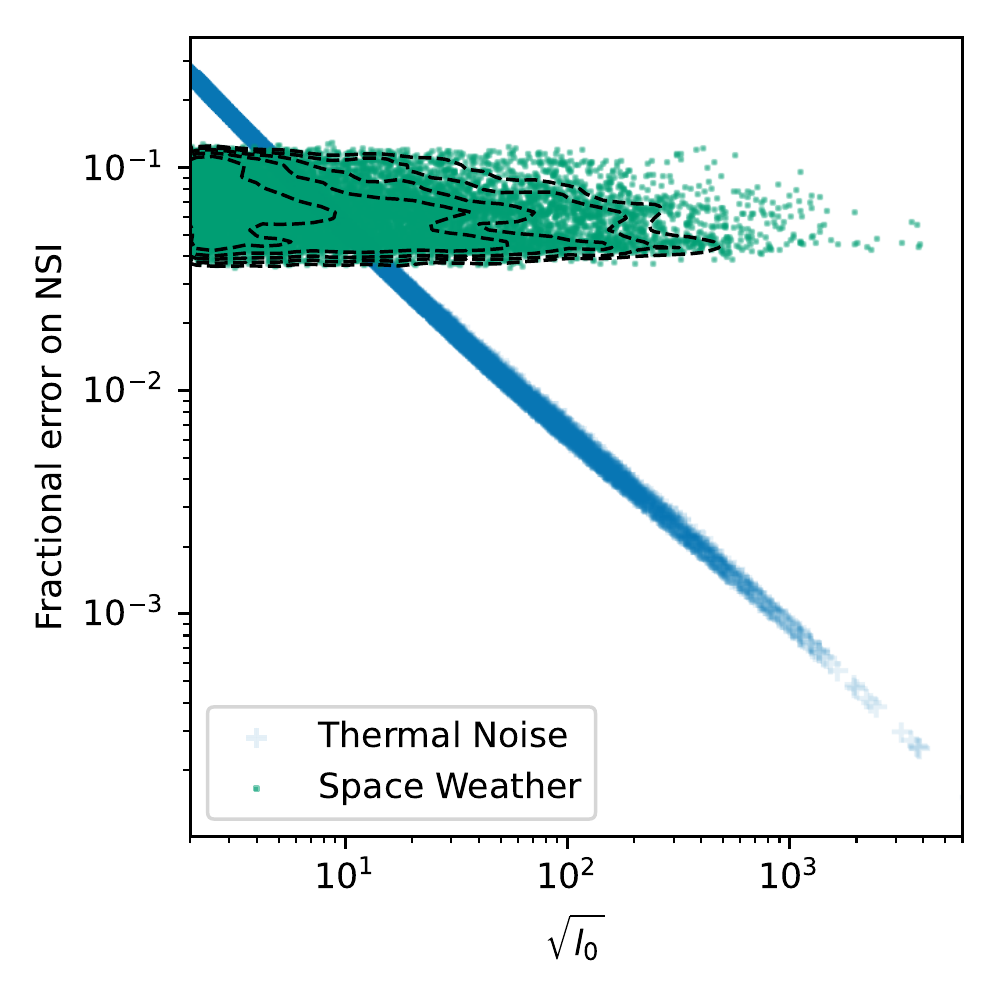}
    \caption{\label{fig:error_budget}The two main sources of error on the NSI (plotted as a fractional error) as a function of S/N. Dashed lines show logarithmically-spaced density contours for the Space Weather fractional errors.}
\end{figure}

These two sources of error are plotted separately in \fig~\ref{fig:error_budget}.
They are combined in quadrature as $\textrm{NSI}_\textrm{err}$, leading to a fractional error distribution which is reasonably uniform in the range 4.8\%--11.4\% (which encompasses 80\% of detected sources).

Consistent with previous studies \citep[e.g.][]{1986P&SS...34...93T}, we found an excess of extreme values compared to a Gaussian distribution.

\subsection{Upper limits}
\label{sec:upper_limits}
Upper limits on the NSI are useful, since setting robust limits on the fraction of the radio source that is sufficiently compact to scintillate is a useful astrophysical tool when interpreting radio source morphology.
In part as a result of the non-linearity between the strength of an IPS detection, and the implied scintillating flux density (\eqn~\ref{eqn:p_to_ds}) we have a very large number of sources for which scintillation is not unambiguously detected, but where we can rule out an NSI of 1.
This is illustrated clearly in the bottom-right panel of \fig~\ref{fig:basic_el}, which shows the residual sum of squares for such a source as a function of NSI. $RSS\left(\textrm{NSI}\right)$ is almost flat below $\textrm{NSI}=0.25$ before increasing more steeply than quadratically for higher values of $\textrm{NSI}$.

We characterise the upper limits via two numbers: $\textrm{UL}_\textrm{fit}$ and $\textrm{UL}_\textrm{err}$.
The former is the maximum-likelihood NSI using a least squares fit (or zero if the fitted NSI$<0$).
$\textrm{UL}_\textrm{err}$ is upper error on $\textrm{UL}_\textrm{fit}$ (allowing for asymmetric errors), determined as being the NSI at which the sum of squares increases by 1 (c.f. \eqn~\ref{eqn:nsi_err}).

While $\textrm{UL}_\textrm{fit}$ and $\textrm{UL}_\textrm{err}$ are derived in a very similar way to $\textrm{NSI}_\textrm{fit}$ and $\textrm{NSI}_\textrm{err}$, it is critical to emphasise that $\textrm{UL}_\textrm{fit}$ is not a useful estimate of the NSI, and any NSI less than $\textrm{UL}_\textrm{fit}$ is consistent with our data.
For this reason we give this column a name that cannot be confused with an NSI by a casual user of our catalogue.

For sources for which upper limits are given, an approximate (un-normalised) log probability distribution function for the NSI given our data is given by
\begin{equation}
	\label{eqn:ul_logpdf}
    \log\left(p\left(\textrm{NSI}\right)\right) = 
    \begin{cases}
    -\left(\frac{\textrm{NSI}-\textrm{UL}_\textrm{fit}}{\textrm{UL}_\textrm{err}}\right)^2 &;\ \textrm{NSI}>\textrm{UL}_\textrm{fit} \\
    \hspace{8ex} 0 &;\ \text{otherwise} \\
    \end{cases}
\end{equation}
i.e. Gaussian in likelihood above the maximum likelihood NSI, uniform below. 

\subsection{Catalogue Contents}
\label{sec:cat}
We now split all of our sample of sources (as defined in \sect~\ref{sec:gleamxmatch}) into the following categories: 
``detected'' sources are those as defined above. 
We further define ``marginal'' detections as those for which $4<l_0\le25$ (i.e. 2-5$\sigma$ detections).
``upper-limit'' sources are sources that are not ``detected'' or ``marginal'', but satisfy the criterion for an informative upper limit ($\textrm{NSI}_\textrm{fit} < 1$; $l_1>9$).
Any remaining sources are not listed in the final catalogue.
$\textrm{NSI}_\textrm{fit}$ and $\textrm{NSI}_\textrm{err}$ are provided only for ``detected'' and ``marginal'' sources; $\textrm{UL}_\textrm{fit}$ and $\textrm{UL}_\textrm{err}$ are only provided for ``upper limit'' sources.

The low threshold of 2$\sigma$ for the most ``marginal'' detections is chosen since these sources have roughly symmetric errors within $\pm 1\sigma$ (see \fig~\ref{fig:basic_el}) and so are best described as a ``marginal'' detection with symmetric errors rather than an ``upper-limit'' with asymmetric errors.
Clearly some will be false detections (even if the majority are true).
In any case, $l_0$ and $l_1$ are provided in the catalogue (allowing, e.g. $<3\sigma$ detections to be excluded or treated as upper limits where appropriate).

The GLEAM ID and the flux density of the GLEAM source at the relevant frequency is also provided. Our observing bandwidth is aligned with GLEAM, and the GLEAM flux density we report for each source is an (inverse variance weighted) average derived from the relevant GLEAM measurements and their errors.
We also list $S_\textrm{5lim}$: the product of $\textrm{NSI}_\textrm{5lim}$ (see \sect~\ref{sec:class}) and the GLEAM flux density. 
This provides an estimate of the detection limit of the scintillating component of a source at that location.
$S_\textrm{cont lim}$, the detection limit in the standard images (see \sect~\ref{sec:sensitivity_cont_var}) is also provided.

These are shown in \tab~\ref{tab:cat}.
A description of all columns in the catalogue is given in \tab~\ref{tab:cols}.
The full table is available online\footnote{\url{https://cloudstor.aarnet.edu.au/plus/s/zhj8DMJwyq3T2zQ}}
\begin{sidewaystable}
\caption{\label{tab:cat} First 11 lines of catalogue. Columns are fully described in \tab~\ref{tab:cols}.}
\footnotesize

\setlength{\tabcolsep}{5pt}
\begin{tabular}{crrccccccccccccc}
    \hspace{0.5ex}
\texttt{GLEAM} & \texttt{RAJ2000} & \texttt{DEJ2000} & \texttt{s\_162} & \texttt{elongation} & \texttt{class} & \texttt{nsi\_fit} & \texttt{nsi\_err} & \texttt{ul\_fit} & \texttt{ul\_err} & l\texttt{\_0} & \texttt{l\_1} & \texttt{s\_cont\_lim} & \texttt{s\_5lim} & \texttt{n\_fit} & \texttt{n\_eff} \\
(1) & (2) & (3) & (4) & (5) & (6) & (7) & (8) & (9) & (10) & (11) & (12) & (13) & (14) & (15) & (16) \\
    \hspace{0.5ex}
J000311+093114 & 0.79639 & +9.52069 & 0.263 & 28.3 & \texttt{upper\_limit} & -- & -- & 0.146 & 0.139 & 0.1 & 217.3 & 0.245 & 0.146 & 5 & 4.35 \\
J000315+111538 & 0.81368 & +11.26067 & 0.535 & 31.3 & \texttt{upper\_limit} & -- & -- & 0.000 & 0.191 & 0.1 & 310.5 & 0.269 & 0.276 & 5 & 4.65 \\
J000327+105553 & 0.86297 & +10.93145 & 0.312 & 30.3 & \texttt{upper\_limit} & -- & -- & 0.313 & 0.082 & 2.9 & 198.2 & 0.177 & 0.174 & 6 & 5.34 \\
J000501+095054 & 1.25791 & +9.84843 & 0.265 & 31.2 & \texttt{upper\_limit} & -- & -- & 0.451 & 0.119 & 2.9 & 39.5 & 0.262 & 0.207 & 5 & 4.35 \\
J000515+102442 & 1.31362 & +10.41180 & 0.249 & 31.7 & \texttt{upper\_limit} & -- & -- & 0.448 & 0.111 & 3.4 & 46.5 & 0.222 & 0.186 & 5 & 4.42 \\
J000556+093701 & 1.48703 & +9.61714 & 0.304 & 30.2 & \texttt{detected} & 0.802 & 0.101 & -- & -- & 78.1 & 20.1 & 0.194 & 0.186 & 6 & 5.14 \\
J000608+105209 & 1.53332 & +10.86923 & 0.352 & 30.3 & \texttt{upper\_limit} & -- & -- & 0.119 & 0.177 & 0.0 & 155.2 & 0.200 & 0.223 & 6 & 5.53 \\
J000634+084808 & 1.64536 & +8.80223 & 0.262 & 30.3 & \texttt{upper\_limit} & -- & -- & 0.131 & 0.172 & 0.1 & 147.4 & 0.195 & 0.167 & 6 & 5.21 \\
J000649+082014 & 1.70647 & +8.33732 & 0.756 & 30.6 & \texttt{detected} & 0.212 & 0.033 & -- & -- & 26.2 & 4926.3 & 0.166 & 0.158 & 5 & 4.67 \\
J000707+093004 & 1.78035 & +9.50130 & 0.274 & 30.2 & \texttt{upper\_limit} & -- & -- & 0.303 & 0.093 & 2.0 & 164.9 & 0.158 & 0.162 & 6 & 5.24 \\
J000712+105434 & 1.80150 & +10.90963 & 0.315 & 30.5 & \texttt{marginal} & 0.596 & 0.098 & -- & -- & 17.4 & 49.2 & 0.182 & 0.212 & 6 & 5.47 \\
    \hspace{0.5ex}
\end{tabular}
\end{sidewaystable}
\begin{table*}
\caption{\label{tab:cols} Description of all columns in catalogue (\tab~\ref{tab:cat}) }
\footnotesize
    \begin{tabular}{rcccl}
        Number & Name & ASCII Name     & Units & Description \\
	1      & --   & \texttt{GLEAM} & --    & GLEAM ID from \citet{2017MNRAS.464.1146H} (searchable in NED if prepended with GLEAM) \\ 
        2      & --   & \texttt{RAJ2000} & deg & GLEAM RA \\
        3      & --   & \texttt{DEJ2000} & deg & GLEAM Decl. \\
        4      & $S_\textrm{162}$   & \texttt{s\_162} & jansky & GLEAM flux density at 162\,MHz (see \sect~\ref{sec:cat}) \\
        5      & $\epsilon$   & \texttt{elongation} & deg & Weighted mean solar elongation of observations \\
        6      & --   & \texttt{class} & -- & Class: `detected', `marginal' or `upper\_limit' (see \sect~\ref{sec:cat}) \\
        7      & $\textrm{NSI}_\textrm{fit}$ & \texttt{nsi\_fit} & -- & Normalised Scintillation Index  (NSI; see \eqn~\ref{eqn:s}) \\
        8      & $\textrm{NSI}_\textrm{err}$ & \texttt{nsi\_err} & -- & Error on NSI (see \sect~\ref{sec:nsi_err}) \\
        9      & $\textrm{UL}_\textrm{fit}$ & \texttt{ul\_fit} & -- & Upper limit on NSI (see \sect~\ref{sec:upper_limits}) \\
        10      & $\textrm{UL}_\textrm{err}$ & \texttt{ul\_err} & -- & Error on Upper limit on NSI (see \sect~\ref{sec:upper_limits}) \\
        11      & $l_0$ & \texttt{l\_0} & -- & Measure of likelihood that NSI is non-zero (see \eqn~\ref{eqn:l_0}) \\
        12      & $l_1$ & \texttt{l\_1} & -- & Measure of likelihood that true NSI is non-unity (see \eqn~\ref{eqn:l_1}) \\
        13      & $S_\textrm{cont lim}$ & \texttt{s\_cont\_lim} & jansky & Continuum limit on detection (see \sects~\ref{sec:class}\&\ref{sec:cat}) \\
        14      & $S_\textrm{5lim}$ & \texttt{s\_5lim} & jansky & Limit on variability detection (see \sects~\ref{sec:class}\&\ref{sec:cat}) \\
        15      & -- & \texttt{n\_fit} & -- & Number of observations used in fit \\
        16      & $N_\textrm{eff}$ & \texttt{n\_eff} & -- & Effective number of observations used in fit taking weights into account (see \eqn~\ref{eqn:n_eff}) \\
    \end{tabular}
\end{table*}

\section{Sensitivity, Reliability and Completeness}
\label{sec:sensitivity}
\begin{figure}
    \centering
    \includegraphics[width=1.0\columnwidth]{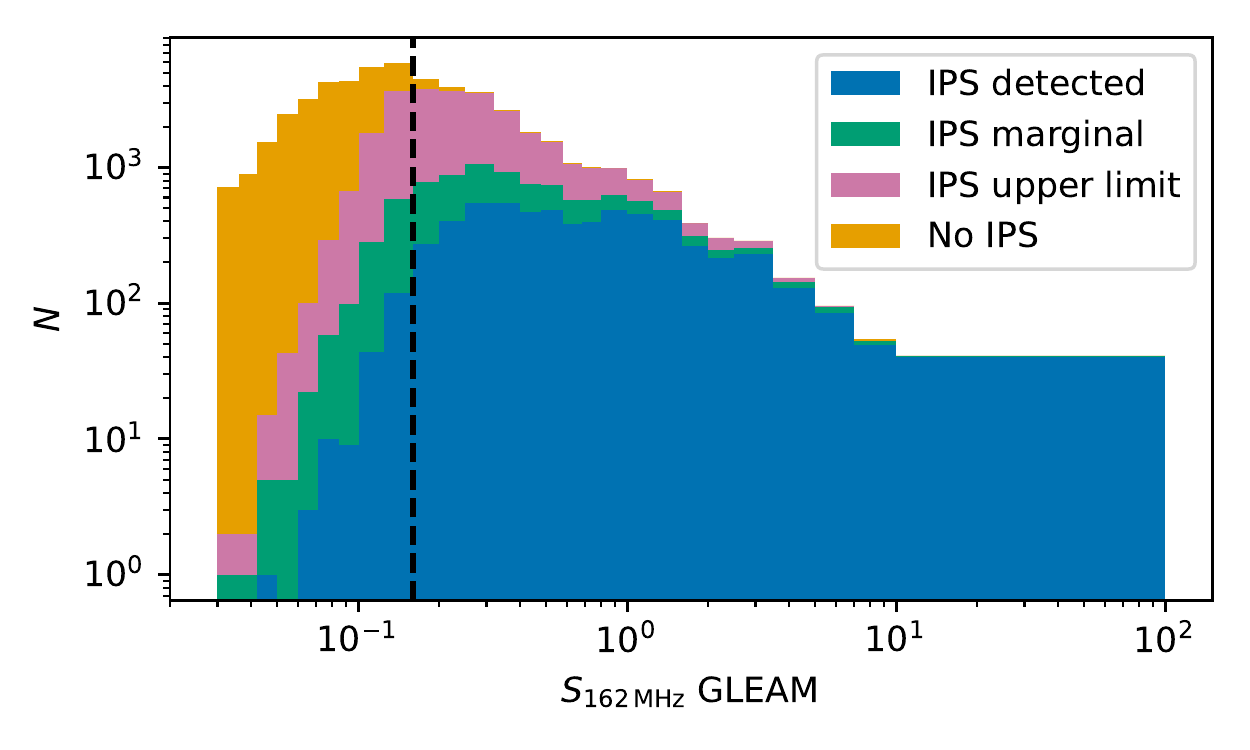}
    \caption{\label{fig:hist_s} Stacked histogram of all GLEAM sources within 1\,hr $<$ RA $<$ 11\,hr; $-10\degr<$ Decl.$<+20\degr$, with colour showing status in our catalogue: detected, marginal, upper limit, not in catalogue. Bins following \citet{2016MNRAS.459.3314F}. Vertical dashed line indicates 0.16\,Jy, the lower edge of the bin in which 82.5\% of sources are in the IPS catalogue.}
\end{figure}

The result of the methodology outlined in the previous two sections is a table of 7839 detections, 5550 marginal detections, and 26367 sources with upper limits.
74\% of these lie within an area of 4875 square degrees bounded by 1\,hr$<$RA $<$11\,hr; $-10<$Decl.$<+20$.
For that region, \fig~\ref{fig:hist_s} shows the sources for each of these categories in the context of all the GLEAM sources in the same region.
Almost all sources brighter than 160\,mJy at our observing frequency appear in our catalogue.
In the remainder of this section we quantify more clearly the sensitivity as a function of survey area, and describe various complications which may affect the reliability and completeness of our survey.

\subsection{Sensitivity}
\label{sec:sensitivity2}
In contrast to most astronomical surveys, our sensitivity at a given location on the sky depends on two factors.
This is because detections must fulfil both the standard image detection criteria, and the variability detection criteria.
While the detection limits imposed by these two sets of criteria are correlated, they are quite distinct: with the former dominated by sidelobe confusion, and the latter dominated by thermal noise (though Sun--source geometry also plays an important role).
Over much of the survey area, the continuum detection threshold is significantly lower than the variability detection threshold, and so, with certain caveats  discussed below, the sensitivity to variability alone determines whether a source is detected or not, and the detection limit for a particular compact flux density is well defined.

In limited sky areas, our variability measurements are more sensitive than the continuum measurements.
This might mean, for example, that a 100\,mJy compact component can be detected if it is a component of a 1\,Jy continuum source (i.e. NSI of 10\%) but not if it is an isolated compact source (i.e. NSI of 100\%).

Another less fundamental issue is the difficulty of assessing our sensitivity for an arbitrary point on the sky.
We sidestep this problem by calculating our sensitivity metrics only at the locations of GLEAM sources which fulfil our selection criteria.
The reader can estimate the sensitivity for an arbitrary location by assuming the sensitivity is the same as it is at the location of nearest catalogued source.
The area of the celestial sphere which is closest to a given source than any other (i.e. the Voronoi cell of each source) has been calculated \citep{Caroli:2010} and these areas are used to determine sensitivity metrics as a function of survey area presented below.

\subsubsection{Sensitivity in continuum and variability}
\label{sec:sensitivity_cont_var}
The key criterion for continuum detection is 5 detections at 5$\sigma$, so the detection threshold for a given source location is simply 5$\times$ the RMS noise in the 5th highest S/N continuum detection $S_\textrm{cont lim}$ (see \sect~\ref{sec:cat}).
The detection threshold for detection of variability (i.e. compact structure) is complicated, as it depends on the precise data, but it can be estimated by $\textrm{S}_\textrm{5lim}$.
\begin{figure*}
    \centering
    \includegraphics[width=1.1\textwidth]{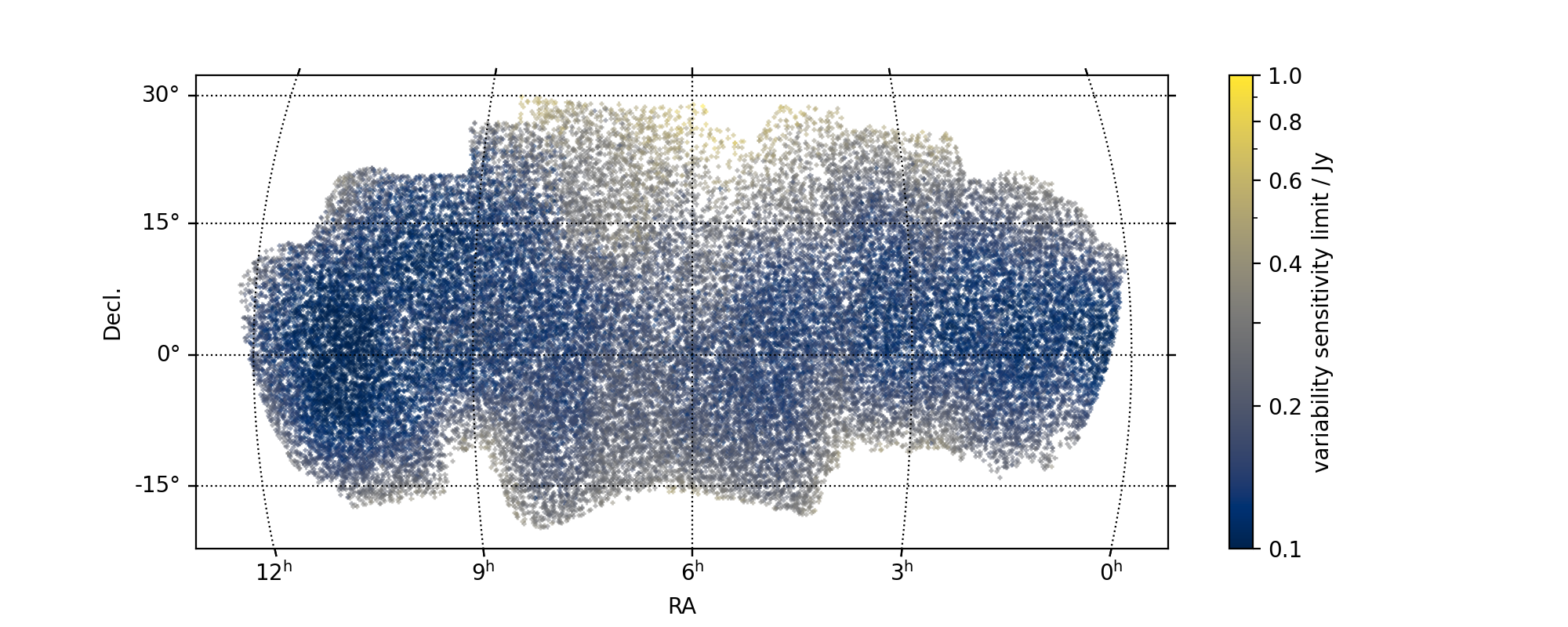}
    \caption{\label{fig:sensitivity_map} Map of all catalogued sources. Colour bar is the compact flux detection limit in jansky.}
\end{figure*}
\Fig~\ref{fig:sensitivity_map} shows the spatial distribution of sensitivity to compact structure. 
The detection limit varies from $\sim$0.1-1\,Jy with good sensitivity being associated with low ecliptic latitude and high Galactic latitude. 

\begin{figure*}
    \centering
    \includegraphics[width=1.0\textwidth]{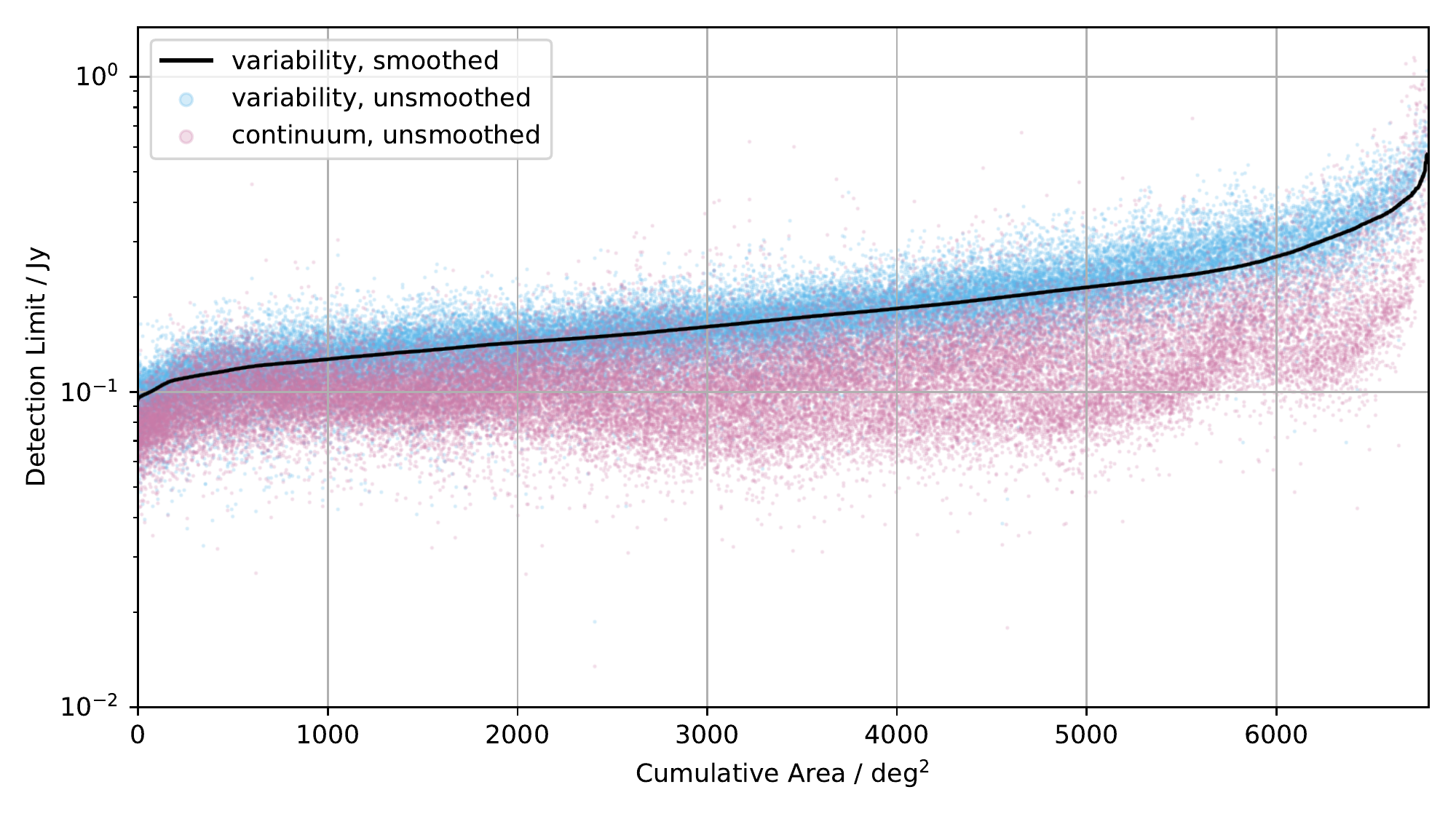}
    \caption{\label{fig:sens_cumulative}Cumulative distribution of (variability) sensitivity as a function of survey area. To construct the black line, the sensitivity measurements at the location of each source (shown in \fig~\ref{fig:sensitivity_map}) are spatially smoothed, and the sources are put in order of smoothed sensitivity. Each point in the survey area is associated with its nearest source (Voronoi tiling) and thus each source has an area associated with it. The x-axis is then the cumulative area represented by these sources. Thus, approximately 5000 square degrees of survey area have a detection limit (to compact structure) $\le$0.2Jy. Blue points are the variability sensitivity measurements without spatial smoothing. Pink points are the (unsmoothed) continuum sensitivity for the same sources.}
\end{figure*}

The cumulative distribution of sensitivity to compact structure as a function of area is presented in \fig~\ref{fig:sens_cumulative}.
To make this more clear, we have spatially smoothed the sensitivity at the location of each source, by making it a weighted average of all sources that lie within a radius of 5$\degr$, where, following \citep{Epanechnikov:1969},  the relative weight is given by
\begin{equation}
    w\left(r\right) = 
    \begin{cases}
    1-\left(\frac{r}{5}\right)^2 &;\ r<5\degr \\
    0 &;\ \text{otherwise} \\
    \end{cases}
\end{equation}
This shows more clearly that the majority of the survey area has a detection limit below 0.2\,Jy with more than 2000 square degrees with a sensitivity $<$0.15\,Jy.

\fig~\ref{fig:sens_cumulative} also shows that for most locations, the continuum sensitivity limit is below the variability limit.
\begin{figure*}
    \centering
    \includegraphics[width=1.2\textwidth]{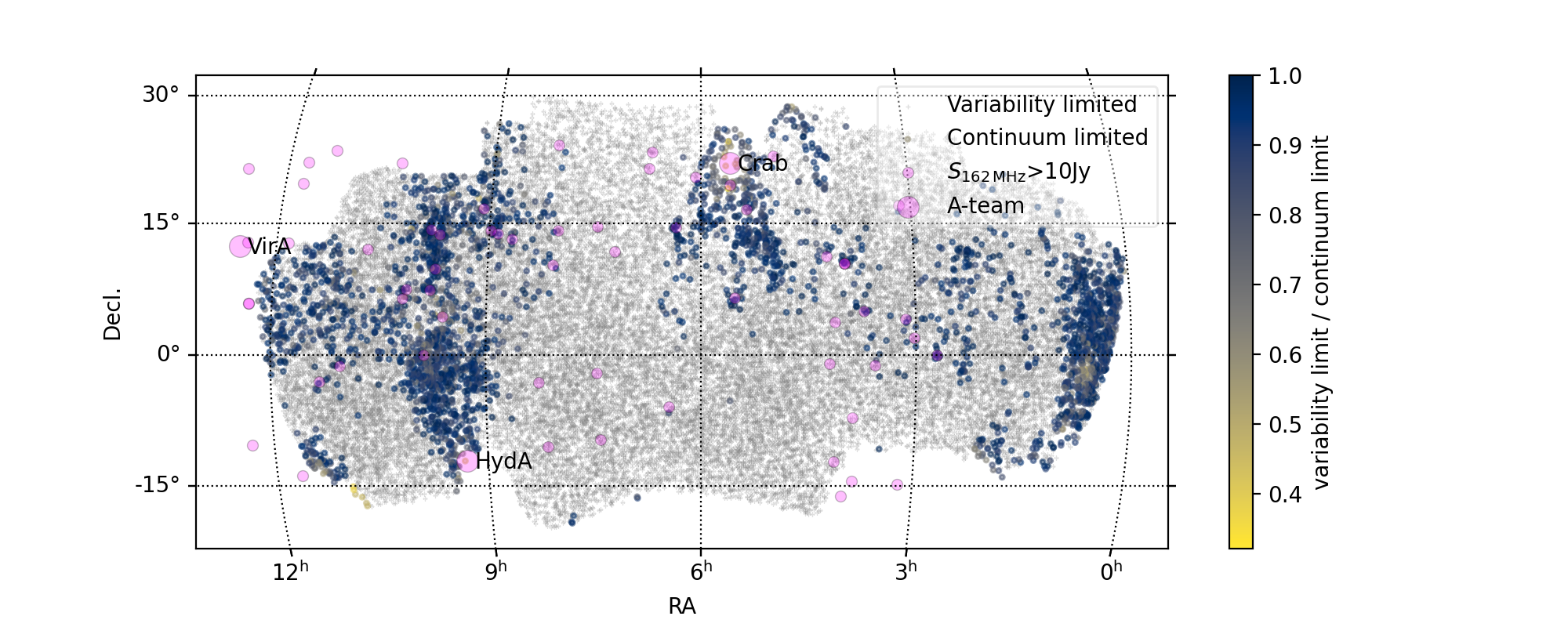}
    \caption{\label{fig:cont_limit}Comparison of variability sensitivity to continuum sensitivity. Grey points (the majority of the map) are those sources where the continuum sensitivity exceeds the variability. Coloured points are those where the variability sensitivity exceeds the continuum sensitivity and the colour bar indicates the ratio. Bright GLEAM sources are shown in magenta, with the size of the points being proportional to the flux density at 162\,MHz.}
\end{figure*}
This is explored further in \fig~\ref{fig:cont_limit} which highlights the rare areas where the inverse is the case.
These areas are associated with very bright continuum sources, since the sidelobes of these sources will cause the noise in the continuum images to be increased.
The edges of the survey area, particularly the Western edge, are also affected since these areas are more sparsely sampled (see \fig~\ref{fig:pointings}).
In many cases, only the bright source itself (and perhaps a small area around it) are affected.
As noted above, in these areas it is possible that a compact source whose IPS variability might have been detected will not make it into the catalogue as it will not have fulfilled the criteria for continuum detection.
However, these areas are quite limited in extent, and only a small fraction of IPS sources are likely to be affected. For example, if the ratio between variability sensitivity and continuum sensitivity is 0.9, then only sources with an NSI $>$0.9 within 10\% of the variability detection limit will be excluded.

There is another mechanism by which source with a lower NSI might be detected while one with a higher NSI (but the same compact flux density) might not.
Sources with a higher NSI may be detected in more continuum images, and therefore will have more variability measurements. 
A number of factors combine to make this effect negligible.
First, most sources in our catalogue are detected in continuum in every observation where the source is within the quarter power point.
Secondly, the impact of additional continuum detections on the IPS sensitivity is modest, both because they will have low weight, and because IPS sensitivity only increases with the fourth power of the number of observations for sources near the detection limit.

\subsection{False detections}
\label{sec:false}
In addition to the 2.5 million 5$\sigma$ detections within 1\arcmin\ of a GLEAM source, there are an additional 854 unassociated 5$\sigma$ detections of variability, within the GLEAM coverage area (and fulfilling the criteria of being within the quarter power point of the primary beam).
Most of these can be explained: the Crab Nebula and Hydra A are deliberately excluded from GLEAM and account for 45 detections, as well as 68 detections which are deemed to be their sidelobes.
10 sources appear to be genuine IPS detections from the Western hotspot of PKS0945+07, a wide double which is characterised in GLEAM (J094746+072509) as an ellipse not fully encompassing the source.
Similarly 3 detections appear to be associated with the Northern component\footnote{NB \citet{2020PASA...37...18W} consider the two components of GLEAM J101051-020137 to be unassociated.} of GLEAM J101051-020137.
17 detections appear to be genuine detections of IPS from TGSSADR J093908.4+020059, a 1\,Jy source (not in GLEAM, but situated approximately 2\arcmin\ from GLEAM J093918+015948, a 4\,Jy source).
The pulsar PSR 0953+10, which is not in GLEAM, is also detected 3 times, the only source below the sensitivity limit of GLEAM from which we have unambiguously detected variability.
The Moon is detected twice (probably due to reflected RFI; \citealp{2013AJ....145...23M}), while Jupiter (which might be expected to give rise to continuum and variability detections) is outside our survey area.

A notable source of false detections is the Pulsar B0950+08 (GLEAM J095309+075539), which is an extremely bright pulsar with a period of $\sim$0.25\,s and has one of the lowest dispersion measures known \citep{1968Natur.218..126P,2005AJ....129.1993M}.
This pulsar produces strong, sidelobe-like artefacts within a radius of approximately 7 degrees, accounting for 116 false detections.
These artefacts are much more pronounced, and more symmetric, than the sidelobes we occasionally see around IPS sources, and therefore are unlikely to be caused by time variability alone.
The highly symmetric nature of these artefacts suggests an amplitude error, and we suggest that it arises due the pulses only occupying a narrow line in the dynamic spectrum, meaning that the pulsar cannot be correctly deconvolved using the PSF for the full bandwidth.

The remaining 500 (presumed) false detections are not evenly distributed.
Just 62 observations account for 301 of the false detections, while 41 observations have no false detections and 83 have just one.
A couple of observations have visible artefacts due to solar bursts, and we presume that most of the remaining observations with 5 or more false detections have similar issues at a much lower level (i.e. either solar activity or RFI).
The number of observations with 0--4 false detections is consistent with a Poisson distribution with a mean of approximately 1.75.
With approximately 737\,000 resolution units within the quarter power point of the primary beam for a typical observation, the appropriate $\chi$ distribution (see \sect~\ref{sec:images}) would predict only 0.34 detections/observation more than 5 standard deviations from the mean.
On the other hand, the rate 1.75 detections/observation is reached at only a slightly lower significance: 4.65$\sigma$.
This underlines the exceptionally clean nature of our data, and is testament to the extremely quiet observing conditions at the Murchison Radio Observatory.
It bodes extremely well for the use of the MWA to detect further transients similar to GLEAM-X J162759.5-523504.3 \citep{2022Natur.601..526H} as well as other radio transients that might be expected on similar timescales to IPS \citep[e.g. prompt emission from gamma-ray bursts:][]{2022PASA...39....3T}.

\subsection{Effect of sub-arcminute structure}
\label{sec:arcmin}
As described above, we have used GLEAM as our reference catalogue, and it matches extremely well in resolution with our continuum images.
In order to allow measurements of scintillating flux density in cases where the scintillation signature is not detectable at a high level of significance, we also make point measurements of the variability measurements at the location of the continuum source (see \sect~\ref{sec:gleamxmatch}).
This approach may lead to an underestimate of the scintillating flux density if the centroid of the source at GLEAM resolution does not match with the centroid of the scintillating emission.

We can measure this effect for sources that have sufficiently high S/N in individual observations that we can determine the offset between continuum and variability centroid.
3920 sources have at least 3 direct detections in the variability image.
Although this sample of ``detected'' sources may be expected to be biased towards more compact objects, the full range of NSIs is represented, and only 1\% of sources show an offset between continuum and variability centroid of more than 35\arcsec. 
Even an offset of this magnitude would only incur a 21\% reduction in the measured scintillating flux density.
However, it is nonetheless possible that some scintillating sources embedded in asymmetrically in extended structure (e.g. single hotspots) may have underestimated NSIs due to this effect.

\subsection{Reliability and completeness}
\label{sec:reliability}
These investigations demonstrate that, for the most part, $S_\textrm{5lim}$ (the proxy for sensitivity, given in the catalogue for the location of each source) gives a good indication of the sensitivity.

Examining the distribution of scatter about the trend-line in \fig~\ref{fig:nsi_snr} in the range $2<\sqrt{l_0}\le10$ we estimate that our ``detected'' sample is 10\% complete at $0.61\textrm{S}_\textrm{5lim}$, 50\% complete at $0.80\textrm{S}_\textrm{5lim}$, 90\% complete at $0.96\textrm{S}_\textrm{5lim}$, and 99\% complete at $1.08\textrm{S}_\textrm{5lim}$. 
The reason that the 50\% completeness is not reached closer to $0.5\textrm{S}_\textrm{5lim}$ is due mainly to the difference between $\textrm{S}_\textrm{lim}$ and $\textrm{S}_\textrm{5lim}$ (see \sect~\ref{sec:class}). 

The extremely small number of false detections of variability give us confidence that our reliability is extremely high, since it would require the chance coincidence of a noise spike with a known GLEAM source on more than one occasion.
The only indication that we have that the reliability is less than 100\% is a small number of sources for which the sum of variability S/N ratios ($\rho_\textrm{var}$; see \sect~\ref{sec:nsi}) is \emph{less than} 0.
This indicates that the variance at the location of the source is less than would be expected due to thermal noise at the location of the source, which is unlikely to be physical. 
A very small number of sources appear to have uniformly negative noise to an extent which is unlikely to be due to chance, and we suggest that this may be due to a poor estimate of the thermal noise level ($\mu$; see \eqn~\ref{eqn:p_to_ds}).
240 such sources have a level of significance that would place them in the ``marginal category'' were the sign of the variability reversed.
None would reach the threshold for the ``detected'' category.

\section{DISCUSSION}
\label{sec:discussion}
\subsection{Comparison with previous work and predictions}
As shown in \fig~\ref{fig:cont_limit}, we achieve a detection limit for a compact component $\le$0.2\,Jy over almost 5000 square degrees.
This is at least a factor of two improvement in sensitivity over our previous work, and is in line with the predicted effect of switching to a more natural visibility weighting scheme \citep[][\sect~3.4]{2019PASA...36....2M}.
The use of data from multiple observations (whether a particular source is detected at 5$\sigma$ in variability or not) has also led to an increase in sensitivity, and it is likely that even greater sensitivity will be realised in the RA range 21\,hr$<$RA$<$24\,hr, where the Galactic latitude is high, but the ecliptic is further south, and therefore closer to the zenith at the MWA.

A fundamental property of our IPS measurements near the detection limit is the asymmetric probability distribution function for the scintillating flux density and parameters derived from it.
This arises since it is necessary to subtract the variance due to scintillation from that due to thermal noise, expressed in \Eqn~\ref{eqn:p_to_ds} \citep[c.f. the ``Debiased Variability Index''][\eqn~1]{2005ApJ...618..108B}.
Since we have 1152 time samples at 0.5\,s resolution (or approximately 500 scintillation timescales) per observation, we are able to detect and measure a scintillation signature even when the variance due to scintillation is only a fraction of that due to thermal noise. 
This means we can robustly detect sources with only approximately 50\,mJy of scintillating flux (corresponding to a compact component of approx. 160\,mJy at 30\degr\ solar elongation).

The unavoidable consequence is that for sources weaker than this, the scintillating flux density cannot be derived from our data, which is consistent with a range of values for the scintillating flux including zero.
However, we derive upper limits on any scintillating flux density for a very large number of sources, ensuring that we extract the maximum amount of information from our data.

\subsection{Future Work}
In this work we have set out in detail the methodology that we will use to conduct IPS surveys.
While there remains some scope for optimisation, we expect that the fundamental approach that we have taken here to combine measurements from multiple observations while keeping track of errors and sensitivity is one that should serve us well into the future as we increase our survey area, and as we move to new instruments, such as ASKAP \citep{2022arXiv220804981C}, and, in time, SKA\_LOW.

The current data release contains more than an order of magnitude more sources than our previous catalogues.
We also probe down to fainter sources and determine the NSI with greater precision.
Therefore, in future work, we can revisit previous topics with greater statistical power \citep{2018MNRAS.474.4937C,2018MNRAS.479.2318C,2019MNRAS.483.1354S}.
We expect that as we are now pushing towards lower flux densities we will be able to provide compactness constraints on sources that are being searched for HI absorption \citep{2020MNRAS.499.4293S}.

We now have IPS measurements for over 250 sources classified by \citet{2017ApJ...836..174C} as having peaked or otherwise curved spectra, over 90\% of which we class as `detected'.
These sources have NSIs ranging from 0.5--1.0 (strongly weighted towards the high end) providing at least some information on angular size, which has been so important in understanding the nature of GHz Peaked Spectrum sources \citep{1998PASP..110..493O}.

The current catalogue also allows us to use the MWA for Space Weather research, by providing a dense network of IPS sources with known scintillation indices.
By comparing the scintillation index observed in an individual MWA observation with this baseline level for each source (the ``g-level''), we are able to map out structures in the solar wind such as CMEs (Morgan et al. in prep.).
In \sect~\ref{sec:nsi_err} we conducted a preliminary analysis of ``g-levels'' with the purpose of determining the error that this imposes on our IPS measurements.
By plotting these ``g-levels'' spatially for each observation, we have already discovered a variety of structures in the solar wind which, due to the space density of our sources, we are able to map in unprecedented detail (Waszewski et al. in prep.).

Another aspect of our data which remains largely unexplored is the wealth of information contained within the timeseries, which in the present work we have boiled down to a single standard deviation measurement.
IPS power spectra are routinely used to characterise the solar wind \citep[e.g.][and references therein]{2021ApJ...922...73T}.
However, weak scintillation power spectra also encode the visibility amplitude of the radio source on scales close to the Fresnel scale \citep{2007MNRAS.380L..20M}.
Sources with an NSI very close to unity are unresolved to IPS, and the power spectrum may not contain source structure information even for IPS-resolved sources (i.e. NSI<1) if the features responsible lie on scales much larger than the Fresnel scale (but sufficiently small that they are unresolved at the 2\arcmin\ interferometric resolution of the MWA; see \sect~2.6 of \citealp{2019PASA...36....2M}).
Preliminary research indicates that data on intermediate scales is responsible for many sources being IPS-resolved, but that substantial structure is also seen close to the Fresnel scale for many sources (Hedge et al. in prep.).

In \sect~\ref{sec:arcmin} we discuss the potential offset of GLEAM source positions from the location of the IPS source (see also \citealp{2018MNRAS.473.2965M} \fig~3 and accompanying text).
The relative locations of the continuum and compact component centroids provides a further quantum of information on the structure of our sources on compact scales.
We defer a more detailed analysis to future work; however we note that approximately half of ``detected'' sources have at least 3 variability detections at 5-$\sigma$, and many of these align well with a particular source component visible in images from TGSS ADR1 \citep{2017A&A...598A..78I}.

Also evident in the current data is that there are areas of the sky with a notable lack of IPS sources.
\begin{figure*}
    \centering
    \includegraphics[width=1.0\textwidth]{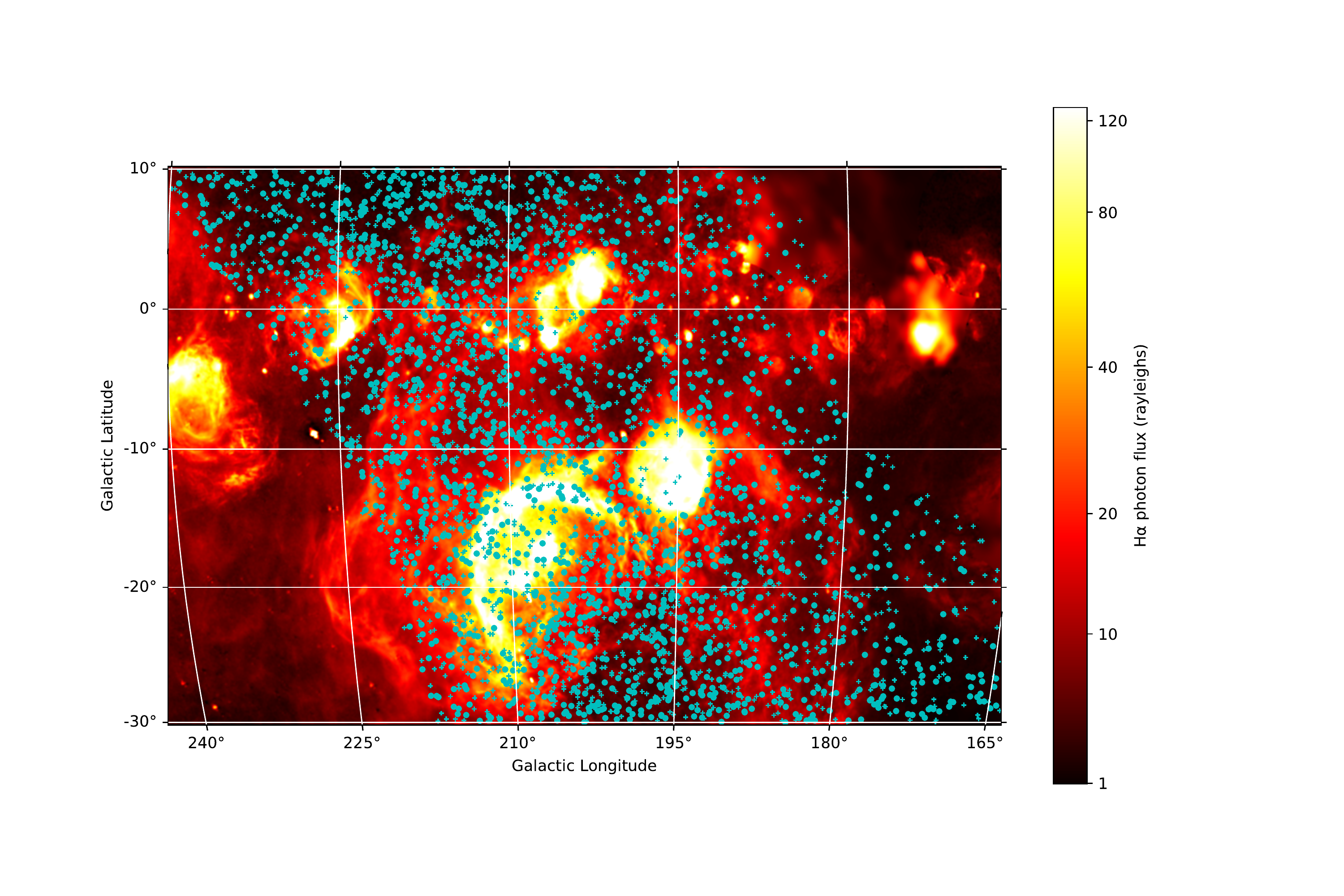}
    \caption{\label{fig:halpha_overlay} Galactic plane (near anti-centre) and Orion region H$\mathrm{\alpha}$ intensity from all-sky image of \citet{2003ApJS..146..407F}  overlaid with our IPS detections (circles) and tentative detections (crosses). Note the very distinct areas where there is a much lower number density of IPS sources, even towards the centre of the survey area. These areas are often associated with H$\mathrm{\alpha}$ emission.}
\end{figure*}
We interpret this as the effect of scatter broadening due to turbulence in the ionised Interstellar Medium (ISM), which has previously been invoked to explain a well-known secular decrease in the number density of IPS sources towards the Galactic plane \citep{1972Natur.236..440R,1984Natur.312..707R,1992Natur.355..232H}.
Now, however, for the very first time, we have sufficient source density to demonstrate not just a trend with galactic latitude, but a very clumpy distribution of ionised turbulence; correlated strongly with H$\mathrm{\alpha}$ emission (see \fig~\ref{fig:halpha_overlay}).
This new science will be explored in a planned publication (Morgan et al. in prep.).



\begin{acknowledgements}
We dedicate this paper to the memory of our friend and colleague Jean-Pierre Macquart.
We thank Natasha Hurley-Walker for making her sky model publicly available.
We thank Natasha Hurley-Walker and Phil Edwards for their careful reading of the manuscript.
This scientific work makes use of the Murchison Radio-astronomy Observatory, operated by CSIRO.
We acknowledge the Wajarri Yamatji people as the traditional owners of the Observatory site.
Support for the operation of the MWA is provided by the Australian Government (NCRIS), under a contract to Curtin University administered by Astronomy Australia Limited.
We acknowledge the Pawsey Supercomputing Centre which is supported by the Western Australian and Australian Governments.
We acknowledge the use of the Legacy Archive for Microwave Background Data Analysis (LAMBDA), part of the High Energy Astrophysics Science Archive Center (HEASARC). HEASARC/LAMBDA is a service of the Astrophysics Science Division at the NASA Goddard Space Flight Center.
\end{acknowledgements}

\bibliographystyle{pasa-mnras}
\bibliography{refs}
\end{document}